\documentclass[authoryear,preprint,review,12pt]{elsarticle}

\usepackage{amssymb}
\def\spose#1{\hbox to 0pt{#1\hss}}
\def\tover#1#2{{\strut\displaystyle #1 \over \strut\displaystyle #2}}

\def\deg{\ifmmode^\circ\else$\null^\circ$\fi}
\def\lta{\mathrel{\spose{\lower 3pt\hbox{$\mathchar "218$}}\raise 2.0pt\hbox{$\mathchar"13C$}}}
\def\gta{\mathrel{\spose{\lower 3pt\hbox{$\mathchar "218$}}\raise 2.0pt\hbox{$\mathchar"13E$}}}
\def\scinot#1.{\hbox{$\, \times \, 10^{#1}$}}

\journal{Icarus}

\begin{document}

\begin{frontmatter}



\title{Seasonal Stratospheric Photochemistry on Uranus and Neptune}


\author[label1]{Julianne I.~Moses}
\author[label2]{Leigh N.~Fletcher}
\author[label3]{Thomas K.~Greathouse}
\author[label4]{Glenn S.~Orton}
\author[label3]{Vincent Hue}

\address[label1]{Space Science Institute, 4750 Walnut Street, Suite 205, Boulder, CO 80301, USA}

\address[label2]{Department of Physics and Astronomy, University of Leicester, University Road, Leicester, LE1 7RH, UK}

\address[label3]{Southwest Research Institute, San Antonio, TX 78228, USA}

\address[label4]{Jet Propulsion Laboratory, MS 183-501, Pasadena, CA 91109, USA}

\begin{abstract}
A time-variable 1D photochemical model is used to study the distribution of stratospheric hydrocarbons as a 
function of altitude, latitude, and season on Uranus and Neptune.  The results for Neptune indicate that in the absence of 
stratospheric circulation or other meridional transport processes, the hydrocarbon abundances exhibit strong seasonal 
and meridional variations in the upper stratosphere, but that these variations become increasingly damped with depth due 
to increasing dynamical and chemical time scales.  At high altitudes, hydrocarbon mixing ratios are typically largest 
where the solar insolation is the greatest, leading to strong hemispheric dichotomies between the summer-to-fall 
hemisphere and winter-to-spring hemisphere.  At mbar pressures and deeper, slower chemistry and diffusion lead to 
latitude variations that become more symmetric about the equator.  On Uranus, the stagnant, poorly mixed stratosphere 
confines methane and its photochemical products to higher pressures, where chemistry and diffusion time scales remain 
large.  Seasonal variations in hydrocarbons are therefore predicted to be more muted on Uranus, despite the planet's 
very large obliquity.  Radiative-transfer simulations demonstrate that latitude variations in hydrocarbons on both 
planets are potentially observable with future JWST mid-infrared spectral imaging.  Our seasonal model predictions 
for Neptune compare well with retrieved C$_2$H$_2$ and C$_2$H$_6$ abundances from spatially resolved ground-based 
observations (no such observations currently exist for Uranus), suggesting that stratospheric circulation --- which 
was not included in these models --- may have little influence on the large-scale meridional hydrocarbon distributions 
on Neptune, unlike the situation on Jupiter and Saturn.
\end{abstract}


\end{frontmatter}

%\linenumbers

\section{Introduction}\label{sec:intro}

Infrared and ultraviolet observations reveal that the stratospheric composition of Uranus and Neptune is being 
altered by solar-driven photochemistry, despite the great distance of these planets from the Sun
\citep[see the reviews of][]{atreya91,bishop95,yung99}.  Methane photolysis by solar ultraviolet radiation triggers 
the production of acetylene (C$_2$H$_2$), ethylene (C$_2$H$_4$), ethane (C$_2$H$_6$), methylacetylene (CH$_3$C$_2$H), 
diacetylene (C$_4$H$_2$), and other complex hydrocarbons, many of which have been observed on Uranus and Neptune \citep[see][and references
therein]{burgdorf06,orton14chem}. These photochemically produced species are radiatively 
active at mid-infrared wavelengths and can affect many aspects of the planetary atmosphere, such as its 
thermal structure, aerosol structure, energy balance, dynamical motions, and ionospheric structure.  A full understanding 
of the three-dimensional (3D) time-variable behavior of photochemically produced species is therefore important for 
understanding many aspects of atmospheric physics and chemistry on Uranus and Neptune.  

% Fig 1, 13 lines
\begin{figure*}[!htb]
%\vspace{-1.5cm}
\begin{center}
\includegraphics[clip=t,width=5.4in]{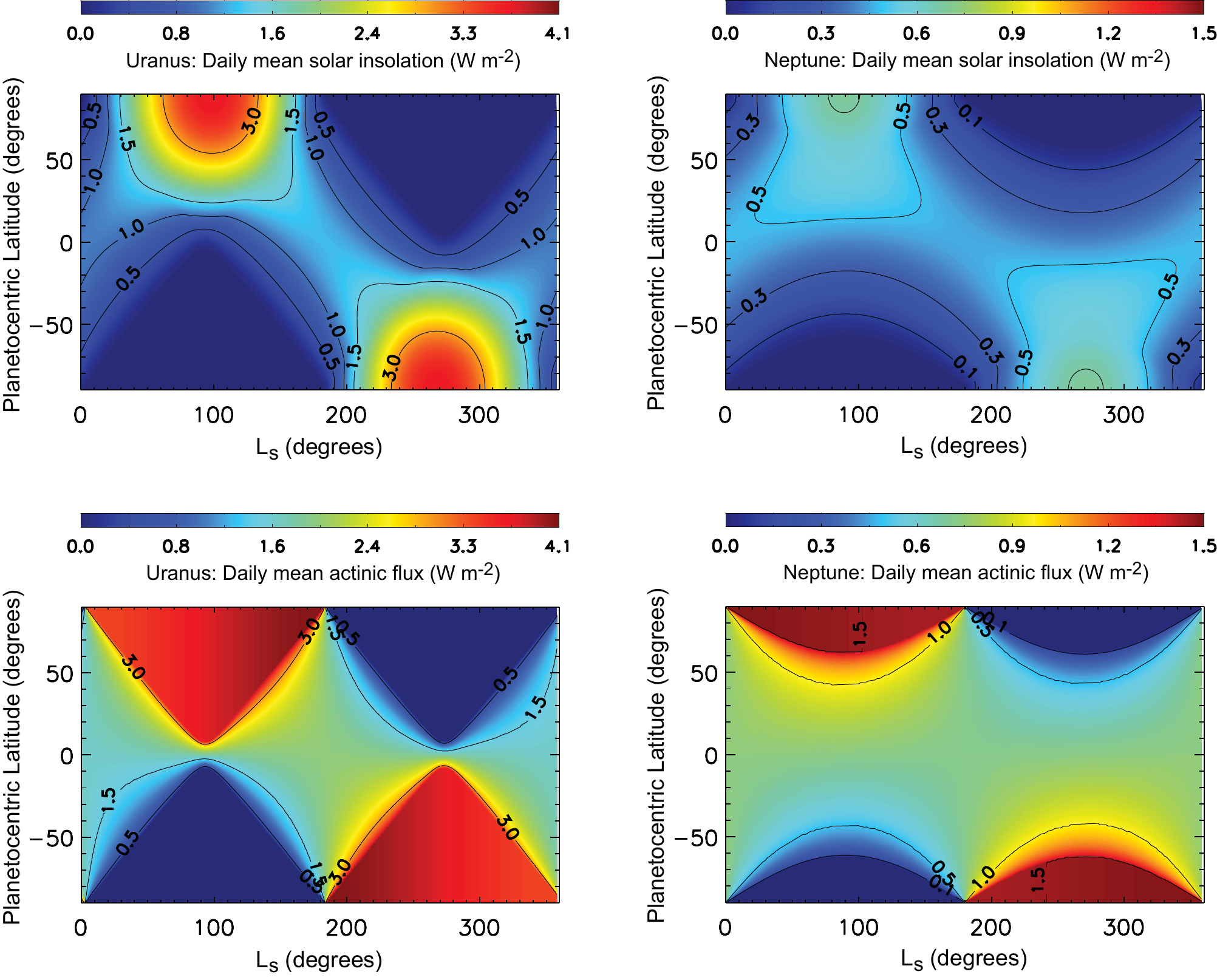}
\end{center}
\vspace{-0.5cm}
\caption{(Top) Mean daily solar insolation (W m$^{-2}$ per planetary day) incident onto a unit horizontal surface at the 
top of the atmosphere of Uranus (Left) and Neptune (Right) as a function of planetocentric latitude and season, 
where season is represented by solar longitude $L_s$.  (Bottom) Mean daily actinic flux (W m$^{-2}$ per planetary day) 
at the top of the atmosphere of Uranus (Left) and Neptune (Right) as a function of planetocentric latitude and season.  Note 
that a molecule being photodissociated does not care what direction the photon is coming from, just what the local photon flux 
is; therefore, the actinic flux, which is the solar flux without accounting for the cosine dependence of the solar zenith 
angle, is more relevant to the photochemistry discussion than the insolation at a ``surface.''\label{figinsol}}
%\vspace{-10pt}
\end{figure*}

\begin{figure*}[!htb]
%\vspace{-1.5cm}
\begin{center}
\includegraphics[clip=t,scale=0.6]{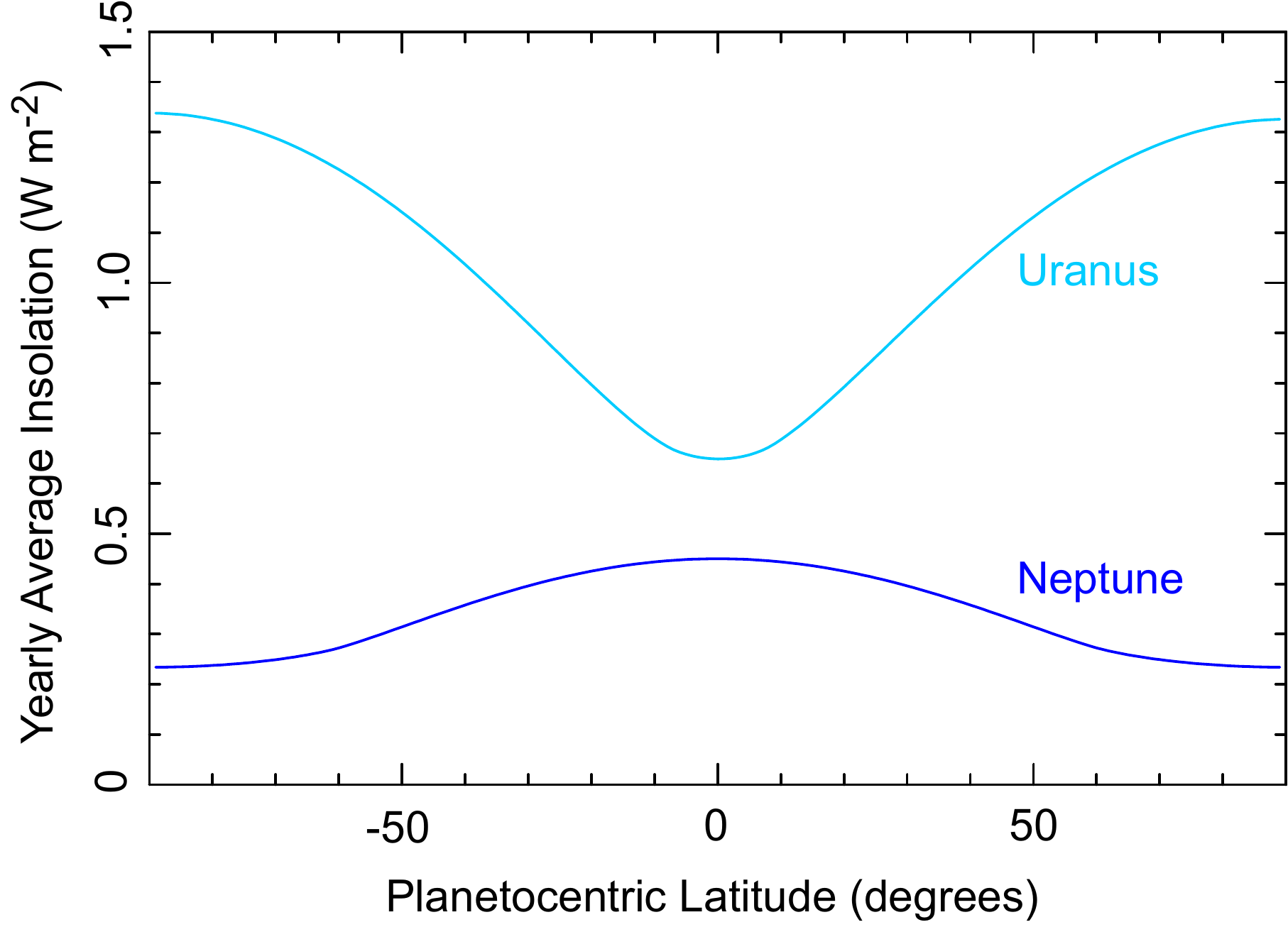}
\end{center}
\vspace{-0.5cm}
\caption{Annual average solar insolation at Uranus and Neptune as a function of latitude.
Unlike the situation on other planets in the solar system, the polar regions of Uranus receive 
a higher annual average insolation than the equatorial region.
\label{figannual}}
%\vspace{-10pt}
\end{figure*}

The non-zero obliquity (axial tilt) of Uranus and Neptune results in a seasonal dependence of solar insolation (see 
Fig.~\ref{figinsol}) that affects the production and loss rates of photochemically active constituents.  Uranus, with its 
extreme $\sim$97.8\deg\ obliquity and rotational pole nearly in line with its orbital plane, experiences very unusual seasons 
compared to other Solar-System planets.  Averaged over a year, high latitudes on Uranus receive greater solar insolation 
than low latitudes (see Fig.~\ref{figannual}).  Much of the planet alternates between being almost fully illuminated 
and being in almost complete darkness for half a year at a time (with one year on Uranus being equal to 84 Earth years), 
creating an opportunity for dramatic changes in hydrocarbon production as a function of season.  
Neptune's more moderate 28.3\deg\ obliquity results in seasonal forcing similar to that on the Earth, Mars, and Saturn, 
with low latitudes receiving a greater annual average solar insolation than high latitudes. Thus, averaged over a year,
hydrocarbon production rates on Neptune will be greater at low latitudes than high latitudes.  
Given that a Neptune year is 165 Earth years, the winter high-latitude regions on Neptune endure long periods of time 
without sunlight, and the reduction in photochemical production of stratospheric hydrocarbons during the long 
polar winter could potentially affect global hydrocarbon abundances and/or result in different meridional distributions 
of hydrocarbons than shorter-period planets with similar obliquities, such as Saturn.  

% Fig 2, 14 lines

Although several one-dimensional (1D) photochemical models 
for Uranus and Neptune have been developed in the past
\citep[e.g.,][]{atreya83,romani88,romani89,romani93,summers89,bishop90,bishop92,bishop98,moses92nucl,moses95c,moses05,lellouch94,dobrijevic98,dobrijevic10,bezard99,schulz99,orton14chem,moses17poppe}, 
all previous models were designed for either global-average conditions or specific latitudes and times.  Here, we present 
results from a 1D time-variable model that tracks the seasonal variation of photochemically produced hydrocarbons as a 
function of altitude for different latitudes.  The models are similar to those of \citet{mosesgr05} in that the 1D 
models for the different latitudes are not connected to each other via atmospheric circulation or any type of 
meridional transport, and the temperature structure is kept constant with latitude and time.  
\citet{hue15} show that for Saturn, the expected seasonal variations in stratospheric 
temperatures have only a minor influence on the abundances of the 
observable hydrocarbons, except in high-latitude regions during winter, where downward diffusion of hydrocarbons is 
faster as a result of atmospheric compression due to the lower temperatures.  On the other hand, atmospheric dynamics can alter 
the vertical and meridional distribution of stratospheric constituents in potentially more significant ways.  Comparisons 
of seasonal 1D photochemical models with the observed vertical and meridional distribution of 
hydrocarbon abundances can therefore provide useful information on the nature and strength of atmospheric dynamics, as 
well as atmospheric chemistry \citep[see the Jupiter and Saturn studies 
of][]{mosesgr05,liang05,nixon07,nixon10,guerlet09,guerlet10,friedson12,sinclair13,sinclair14,zhang13bench,sylvestre15,fletcher15satpole,fletcher16jup,hue15,hue16,moses15}.

We describe our photochemical models in section~\ref{sec:model}, present and discuss the results for the seasonal and 
meridional variations in hydrocarbons on Uranus and Neptune in section~\ref{sec:results}, compare the theoretical predictions 
with available observations in section~\ref{sec:compare}, and discuss implications for future observations and modeling in 
section~\ref{sec:implications}. 

\section{Photochemical model}\label{sec:model}

\begin{figure*}[!htb]
%\vspace{-1.5cm}
\begin{center}
\includegraphics[clip=t,scale=0.5]{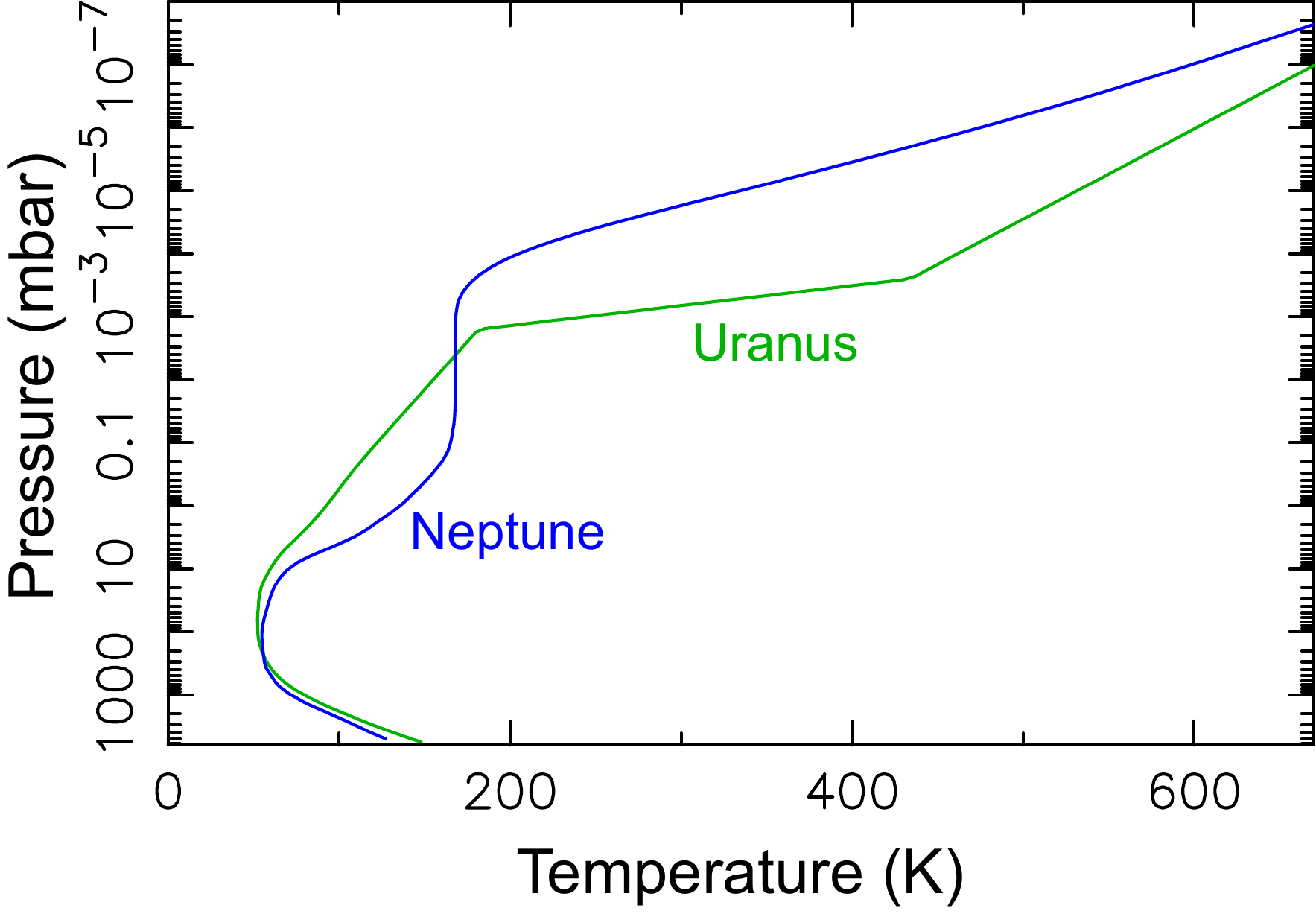}
\end{center}
\vspace{-0.5cm}
\caption{Temperature profiles for Uranus and Neptune (as labeled) adopted in the model.
\label{figtemp}}
%\vspace{-10pt}
\end{figure*}

The coupled set of continuity equations describing the vertical distribution of chemical constituents
in the upper atmosphere of Uranus and Neptune is solved simultaneously by finite-difference methods, using 
the Caltech/JPL 1D KINETICS model \citep[e.g.,][]{allen81,yung84}.  The Uranus model contains 181 vertical 
grid points, ranging from 5.6 bar to 1$\, \times \, 10^{-8}$ mbar, with at least three grid points per scale height to 
ensure accuracy; the 
background temperature structure is taken from \citet{orton14temp}. The Neptune model contains 
198 vertical grid points ranging from 5.0 bar to 1$\, \times \, 10^{-8}$ mbar, with the background temperature 
structure taken from \citet{moses05}, which was originally based on \citet{orton92}, \citet{roques94}, and 
\cite{wangyelle92}.  There is no physically meaningful reason for the different number of grid points for each planet 
--- the numbers were chosen simply to conform to previous models for these planets \citep{orton14chem,moses17poppe}.  
Figure~\ref{figtemp} shows the model temperature grid for both planets.  
Although the temperature-pressure profile is assumed to be constant with 
latitude, the altitude grid changes with latitude due to the variation of gravity with latitude and radius 
on a rapidly spinning, oblate planet \citep[e.g.,][]{lindal85}.  Thirty different latitudes are considered, 
ranging from $-87\deg$ to 87\deg\ planetocentric latitude, every 6 degrees.

% Fig 3, 12 lines

\subsection{Chemistry inputs and boundary conditions}\label{sec:boundary}

The chemical-kinetics inputs to the model are described in \citet{moses05,moses15} and references therein, with 
the complete updated reaction list presented in \citet{moses17poppe}.  A total of $\sim$70 hydrocarbon and oxygen species 
are considered, interacting through $\sim$500 chemical reactions, with 113 of these being photolysis reactions.  The 
model includes condensation of C$_2$H$_2$, C$_2$H$_6$, C$_3$H$_8$, C$_4$H$_2$, C$_4$H$_{10}$, C$_6$H$_6$, H$_2$O, 
and CO$_2$, all of which will condense on both Uranus and Neptune \citep[for the condensation procedure, see][]{moses00b}.  
Multiple Rayleigh scattering of the dominant gas-phase 
constituents is included, but aerosol scattering and absorption are not --- the stratospheric aerosols are optically 
thin \citep{pollack87,moses95c,karkoschka09,karkoschka11}.  The chemical production and loss rates and atmospheric 
transmission profiles are diurnally averaged, which is appropriate given that photolysis time scales and chemical time 
constants for the hydrocarbons are longer than a planetary day ($\sim$17.24 hr rotation rate for Uranus,
$\sim$16.11 hr for Neptune).

Methane from the deep atmosphere is the ultimate source of all carbon-bearing species in the model.  Although 
methane condenses in the upper troposphere of Uranus and Neptune, causing it to be depleted at higher altitudes, some 
CH$_4$ vapor makes it past the tropopause cold trap and into the stratosphere, just as water vapor does on the Earth.  
To streamline the current calculations, we neglect tropospheric methane condensation and instead fix the CH$_4$ 
volume mixing ratio at the lower boundary of the model to be representative of the value observed in the lower 
stratosphere --- 1.6$\scinot-5.$ for Uranus \citep[based on][]{orton14chem} and 1.2$\scinot-3.$ for Neptune 
\citep[based on][]{fletcher10akari,lellouch10,lellouch15}.  
%The larger stratospheric 
%methane abundance on Neptune may be caused by stronger tropospheric convective motions and wave activity in 
%the atmosphere of Neptune as compared with Uranus.  
This assumption leads to an inaccurate methane profile in the troposphere and therefore inaccurate tropospheric hydrocarbon 
chemistry (which is relatively unimportant at these deeper altitudes, in any case), so the tropospheric results are ignored 
throughout the paper; only the stratospheric results are discussed.  We also fix the mixing ratios of helium and CO at 
the lower boundary of the model.  On Uranus, we assume the lower-boundary volume mixing ratio to be 15\% for helium 
\citep{lindal87} and 1.0$\scinot-10.$ for CO \citep[i.e., well below the upper limit of][]{teanby13}.  
On Neptune, we assume the lower-boundary volume mixing ratio to be 19\% for helium \citep{conrath91helium} and 
8$\scinot-8.$ \citep{luszczcook13} for CO.  The rest of the hydrocarbon and oxygen species are assumed to have concentration 
gradients of zero at the lower boundary, allowing the species to flow freely from their source regions at higher altitudes 
through the lower boundary at a maximum possible rate.

For the models presented here, we assume that the CH$_4$ mixing ratio at the base of the stratosphere is 
constant with latitude.  \citet{karkoschka09,karkoschka11}, \citet{sromovsky11,sromovsky14}, \citet{irwin12}, 
\citet{tice13}, \citet{dekleer15}, and \citet{luszczcook16} provide observational evidence that the methane 
abundance in the troposphere varies with latitude on both Uranus and Neptune, with the polar regions being 
depleted in methane with respect to equatorial regions.  The high-latitude depletion in CH$_4$ is particularly 
strong for Uranus.  As discussed in the above references, these variations most likely arise from large-scale 
circulation patterns affecting the methane abundance in its condensation region in the upper troposphere, with 
CH$_4$ enhancements occurring in upwelling regions and depletions in downwelling regions, but the full details 
of this process have yet to be worked out. Moreover, it is currently unclear whether the latitude dependence 
extends into the stratosphere or not, or even how the methane is injected into the stratosphere in the first place.  
Does methane ``leak'' out of the troposphere into the stratosphere preferentially in regions where the 
tropopause is warmer and the local saturation vapor pressure higher \citep[e.g.,][]{orton07nep}, or is the methane 
dynamically  injected into the stratosphere preferentially in regions with strong convective uplift 
\citep[e.g.,][]{karkoschka11,sromovsky14,depater14}?  Given the lack of current information on the stratospheric 
behavior of methane as a function of latitude, our constant-with-latitude assumption provides a first-order look 
at the problem that can be altered when more data become available, such as with future observations from the 
\textit{James Webb Space Telescope} (JWST) \citep{norwood16}.  We do note, however, that when \citet{greathouse11} 
assumed a larger stratospheric methane abundance at the equator relative to the poles at Neptune, the temperature 
predictions from their radiative seasonal models were in closer agreement to the observed temperatures, suggesting 
a latitude dependence of CH$_4$ in the stratosphere of Neptune, at least, is possible.

The oxygen-bearing species in our model are assumed to derive from external sources such as comets 
\citep{lellouch05,hesman07,luszczcook13} or interplanetary dust (\citealt{moses92abl}, \citealt{poppe16}; see also 
\citealt{cavalie14} \& \citealt{moses17poppe}), with CO having a possible significant internal source from the deeper 
troposphere \citep{fegley86,lodders94,fletcher10akari,luszczcook13,cavalie17}.  Other major oxygen species from 
the interior, such as 
H$_2$O and CO$_2$, will condense at depth in the troposphere, with negligible amounts being transported into the 
stratosphere.  To account for the external source of oxygen, we assume that H$_2$O, CO, and CO$_2$ are flowing 
in at the top of the atmosphere on both planets.  The relative fluxes for these species are taken from 
\citet{moses17poppe}, who investigated the coupled oxygen-hydrocarbon photochemistry on Uranus and Neptune, 
and constrained the influx rates through comparisons of models with stratospheric observations of 
H$_2$O \citep{feuchtgruber97,feuchtgruber99,lellouch10}, 
CO \citep{lellouch05,hesman07,fletcher10akari,luszczcook13,teanby13,cavalie14}, and 
CO$_2$ \citep{feuchtgruber99,meadows08,orton14chem}.  For Uranus, these influx rates are 1.2$\scinot5.$ H$_2$O 
molecules cm$^{-2}$ s$^{-1}$, 2.7$\scinot5.$ CO molecules cm$^{-2}$ s$^{-1}$, and 2.8$\scinot3.$ CO$_2$ 
molecules cm$^{-2}$ s$^{-1}$ at the top boundary.  For Neptune, the assumed influx rates are 2.0$\scinot5.$ H$_2$O molecules 
cm$^{-2}$ s$^{-1}$, 2.0$\scinot8.$ CO molecules cm$^{-2}$ s$^{-1}$, and 2.3$\scinot4.$ CO$_2$ molecules 
cm$^{-2}$ s$^{-1}$ at the top boundary.  

Note that our assumption of a constant influx at the top boundary provides 
a reasonable approximation to the situation in which the oxygen is supplied by interplanetary dust particles, but 
the vertical profiles of the oxygen species would be more complicated if the bulk of the oxygen were supplied by a 
relatively recent large cometary impact, as has been suggested for Neptune in particular
\citep[see][]{lellouch05,hesman07,luszczcook13,moses17poppe}.  We also assume a downward flux of atomic H at the 
upper boundary of both planets to account for an additional thermospheric source.  We assume an influx rate of 
5$\scinot7.$ and 1$\scinot7.$ H atoms cm$^{-2}$ s$^{-1}$, respectively, at the top boundary of our Uranus and Neptune 
models based on \citet{moses05}.  All other species are assumed to have zero-flux boundary conditions at the top of 
the atmosphere.

\subsection{Vertical transport and solar flux}\label{sec:transport}

\begin{figure*}[!htb]
%\vspace{-1.5cm}
\begin{center}
\includegraphics[clip=t,scale=0.5]{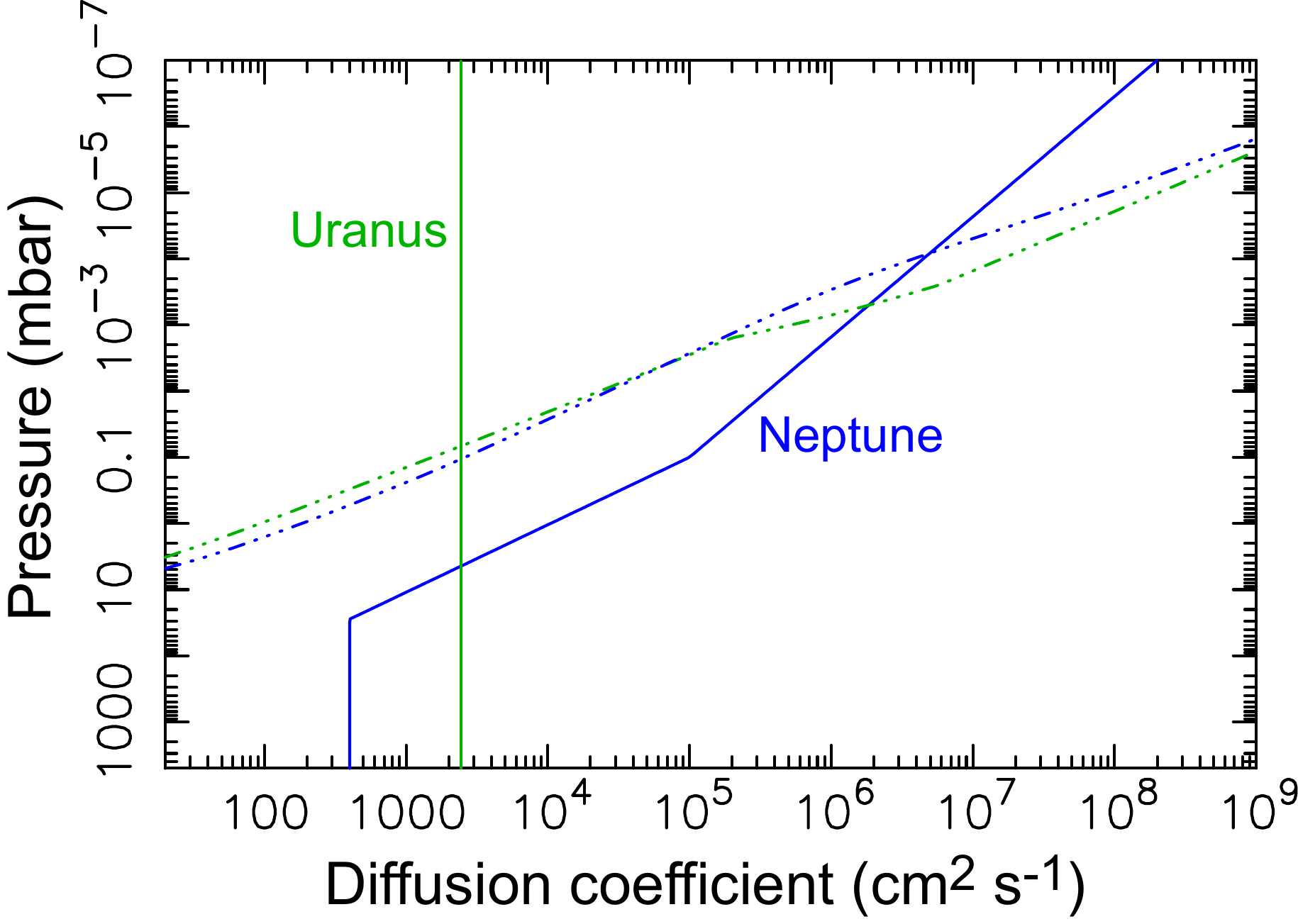}
\end{center}
\vspace{-0.5cm}
\caption{Profiles of the eddy diffusion coefficients (solid lines) and CH$_4$ molecular diffusion coefficients 
(dot-dashed lines) for Uranus (green) and Neptune (blue) adopted in the models.
\label{figeddy}}
%\vspace{-10pt}
\end{figure*}

As is typical in 1D photochemical models, vertical transport is assumed to operate through eddy and molecular diffusion.  
The eddy diffusion coefficient profile is one of the major free parameters for such models.  Figure~\ref{figeddy} shows
the diffusion coefficients adopted in our models. 
For Uranus, we assume the vertical eddy diffusion coefficient $K_{zz} = 2430$ cm$^{2}$ s$^{-1}$, independent of altitude, 
based on \citet{orton14chem}.  That assumption places the methane homopause at the $\sim$7$\scinot-2.$ mbar pressure level.
For Neptune, we assume
\begin{eqnarray}
K_{zz} \, & \, = \, & \, 10^5 \left( \tover{0.1}{P} \right) ^{0.55},\ \textrm{for}\ P\ < \ 0.1\ \textrm{mbar} \nonumber \\
 \        & \, = \, & \, 10^5 \left( \tover{0.1}{P} \right) ^{0.98},\ \textrm{for}\ 28\ \textrm{mbar}\ < \ P\ < \ 0.1\ \textrm{mbar} \nonumber \\
 \        & \, = \, & \, 400,\ \textrm{for}\ P\ > \ 28\ \textrm{mbar}, \nonumber \\
\end{eqnarray}
for $K_{zz}$ in cm$^2$ s$^{-1}$ and $P$ in mbar, which places the methane homopause level at $\sim$8$\scinot-5.$ mbar.  
These eddy diffusion coefficient profiles were chosen in combination with 
our adopted lower-stratospheric CH$_4$ mixing ratio and our updated giant-planet chemical reaction list to provide a reasonable 
fit to the global-average stratospheric hydrocarbon abundances on both Uranus and Neptune 
\citep[see][]{orton14chem,moses17poppe}.  
%For Uranus, the fit to the \textit{Spitzer Space Telescope} Infrared Spectrometer 
%emission data for C$_2$H$_2$ and C$_2$H$_6$ is not as good as the nominal model presented in \citet{orton14chem}. However, 
%the photochemical model developed for that paper used a reaction list where some uncertain reaction rate coefficients were 
%optimized to produce the highest possible C$_2$H$_2$/C$_2$H$_6$ ratio.  This previous reaction list does not provide as good 
%a fit to hydrocarbon abundances on the other giant planets, and our current list provides a compromise that produces a 
%reasonable C$_2$H$_2$/C$_2$H$_6$ ratio for all the giant planets.
For the nominal models presented here, the $K_{zz}$ profile is assumed to be independent of latitude.
The assumptions regarding the molecular diffusion coefficients are described in \citet{moses00a}.  

% Fig 4, 13 lines

For the solar ultraviolet flux, we take the average of the solar-cycle minimum and maximum fluxes presented in 
\citet{woods02}.  Although the hydrocarbon abundances at high altitudes are sensitive to the instantaneous solar 
flux, such that the solar-cycle variations should show up readily in the mixing ratio profiles in the upper 
stratosphere \citep[e.g.,][]{mosesgr05}, chemical and transport time scales are much longer in the middle and 
lower stratosphere where the infrared observations are probing. Therefore, average solar flux values are sufficient 
for our purposes.  We also include an isotropic source of stellar background UV radiation \citep[taken from][]{mathis83} 
and solar Lyman $\alpha$ photons that are being scattered 
from atomic hydrogen in the local interplanetary medium (LIPM).  This LIPM source of photolyzing radiation is increasingly 
important to the photochemistry the farther the planet is located from the Sun, as was first demonstrated by 
\citet{strobel90} and \citet{bishop92}.  The LIPM source is needed particularly for the 
seasonal models presented here, as the scattered Lyman alpha provides a mechanism for CH$_4$ to be photolyzed 
even during the long polar winter on these planets, when high latitudes do not receive direct sunlight.  

The magnitude of this LIPM source depends on both the solar cycle and heliocentric distance.  Before the \textit{Voyager} 
encounter with Neptune, the Ultraviolet Spectrometer (UVS) instrument recorded a background Lyman alpha LIPM intensity 
of roughly 340 Rayleighs \citep{broadfoot89}.  Due to apparent calibration issues with the Voyager measurement 
\citep{gangopadhyay05}, we revise this value downward by a factor of $\sim$2, and based on modeling and observations 
\citep{gangopadhyay05,quemerais09}, we assume the intensity at Uranus is a factor of $\sim$1.3 larger than that at 
Neptune.  The actinitc flux from this LIPM source (assuming isotropic from the upper hemisphere only) is $2 \pi 
\int_0^1 I(\tau, \mu) d\mu$, where $I$ is the intensity incident on a small surface area with cosine of the incidence 
angle $\mu$ at vertical optical depth $\tau$ within the atmosphere.  From Beer's law, $I(\tau, \mu) = I_0 \exp(-\tau/\mu)$, 
and the actinic flux becomes
\begin{equation}
F(\tau) \, = \, 2\pi I_0 \left( e^{-\tau} - \tau \left[ -\textrm{Ei}(-\tau) \right] \right),
\end{equation}
where Ei is the exponential integral.  In general, Lyman alpha photons from this LIPM source are absorbed at 
higher altitudes than the direct solar Lyman alpha source.

Unlike the case on Saturn, the small, faint ring systems of Uranus and Neptune do not cast sufficient shadows on the 
planets to notably affect the solar flux received and so are ignored in the calculations.

\subsection{Seasonal model procedure}\label{sec:procedure}

We first keep the season fixed at northern vernal equinox conditions ($L_s$ = 0\deg) and run the 1D models for the different 
latitudes until steady-state solutions are achieved for that constant season.  Then, we run the time-variable seasonal models 
with the above vernal equinox results as our initial conditions.  The heliocentric distance and relevant solar zenith angle 
and flux calculations 
are updated every planetary day during the seasonal model runs.  The planetary orbital positions as a function of time (and 
$L_s$) for two full planetary years are obtained from the JPL Horizons ephemeris calculator \citep{giorgini96}.  
The values for Uranus begin at $L_s$ = 0\deg\ in November, 1839 and end two planetary years later at $L_s$ = 0\deg\ 
in December, 2007; the values for Neptune begin at $L_s$ = 0\deg\ in September, 1716 and end two planetary years later 
at $L_s$ = 0\deg\ in March, 2046. 
The full two-year time period was chosen originally to avoid any potential discontinuities due to the 
solar cycle being different at the beginning and end of the most recent planetary year when repeating the yearly 
calculations, but with the choice of solar-cycle averages, the full two-year sequence is unnecessary.  By the same 
token, the results for a specific $L_s$ for Neptune from this time period are relevant to the same $L_s$ in the more recent 
era beyond 2007.  Because the chemical species that are produced in the upper stratosphere can take thousands of years to 
diffuse down to the base of the model atmosphere, this two-planetary-year sequence is run again and again, with the results 
from the end of the second year being fed as initial conditions to the beginning of new two-year calculation, until a 
repeatable annual result is obtained.  The Uranus model was iterated for a total of 80 Uranus years ($\sim$6721 Earth years), 
and the Neptune model for 10 Neptune years ($\sim$1648 Earth years) --- more time is needed for the Uranus run because 
of the low eddy diffusion coefficient.  The results from the converged solutions are presented below.

\section{Results and Discussion}\label{sec:results}

The photochemical product abundances vary with altitude, latitude, and time on Uranus and Neptune due to seasonal 
forcing (see full model output in the Supplementary Material).  We begin by discussing the seasonal variations on 
Neptune, which are reasonably straightforward and qualitatively 
similar to those derived for Saturn \citep{mosesgr05,hue15}.  We then discuss seasonal variations on Uranus, which 
differ from those on Neptune and Saturn --- not only because of the planet's extreme axial tilt, but because of the very weak 
vertical mixing on Uranus.  The results for both planets as a function of latitude are then presented, followed by 
the column densities as a function of both latitude and season.
%, followed by a 
%demonstration of the sensitivity of the results to the inclusion of Lyman alpha solar radiation scattered by the LIPM.

% Fig 5, 17 lines
\begin{figure*}[!htb]
%\vspace{-1.5cm}
\begin{center}
\includegraphics[clip=t,width=6.0in]{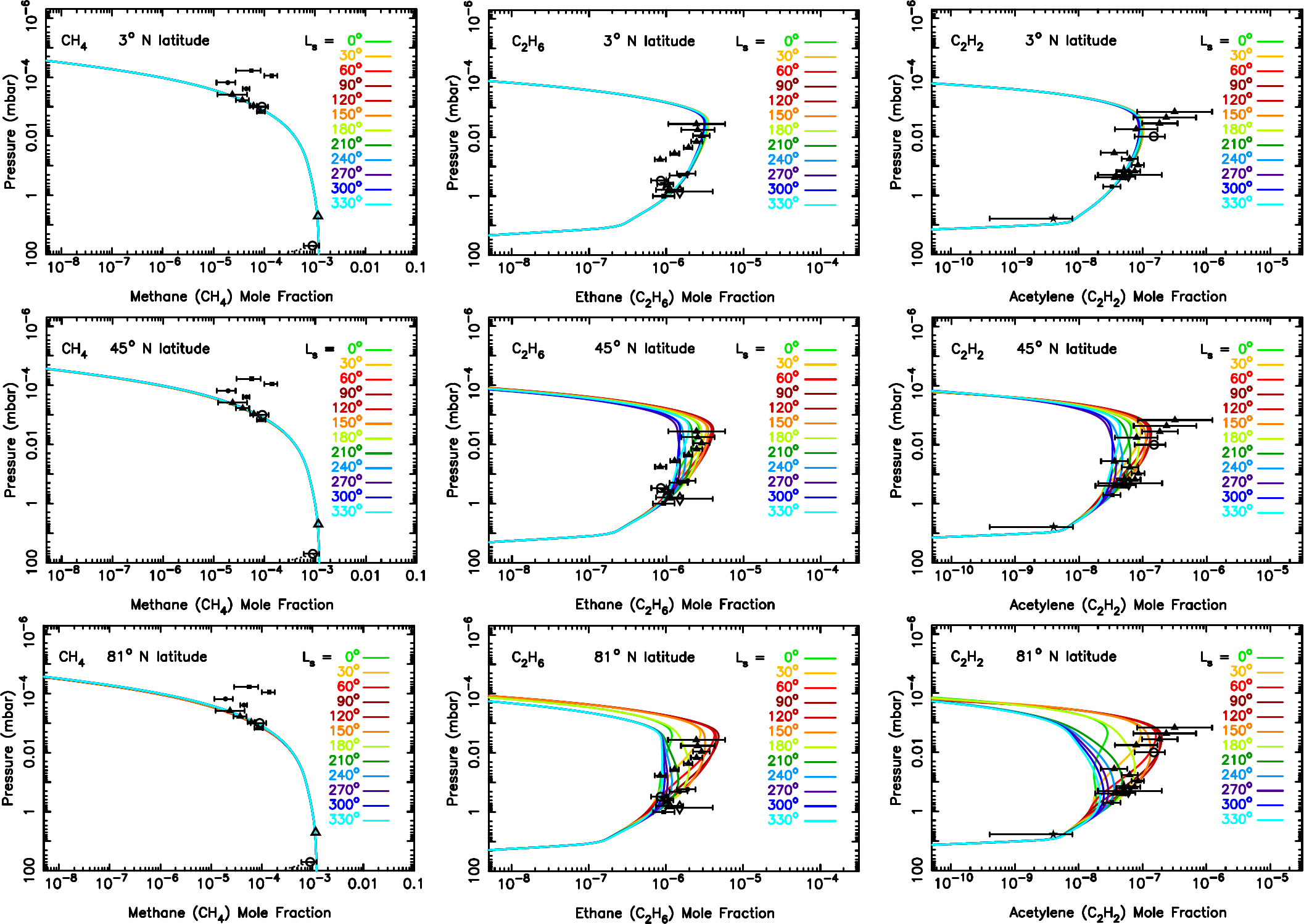}
\end{center}
\vspace{-0.5cm}
\caption{Mixing ratio profiles for key hydrocarbons (as labeled) on Neptune as 
a function of season for 3\deg N planetocentric latitude (Top), 45\deg\ planetocentric latitude (Middle), 
and 81\deg\ planetocentric latitude (Bottom).  Seasons are shown every 30\deg\ in $L_s$.  
Data points are from various published observations (global average or low latitude, acquired during $L_s$ = 230--300\deg)
\citep{caldwell88,bezard91,bishop92,orton92,kostiuk92,yelle93,bezard98,schulz99,meadows08,fletcher10akari,greathouse11,lellouch15},
and unpublished Infrared Space Observatory hydrocarbon observations (Bruno B\'ezard, personal communication, 2001).
\label{figmixnepall}}
%\vspace{-10pt}
\end{figure*}

\clearpage

% Fig 5 (cont), 13 lines
\begin{figure*}[!htb]
%\vspace{-1.5cm}
\begin{center}
\includegraphics[clip=t,width=6.0in]{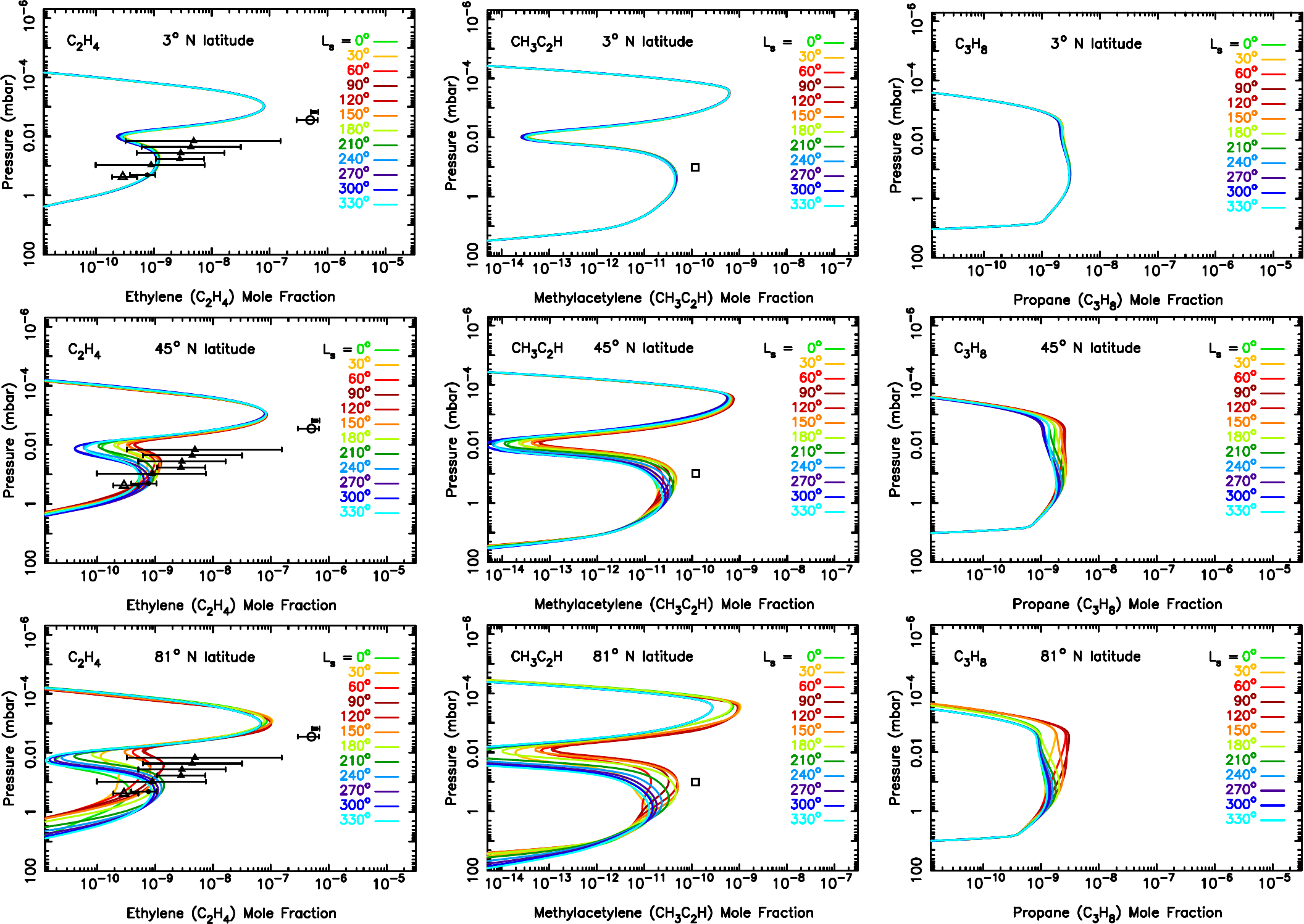}
\end{center}
\vspace{-0.5cm}
\noindent{Fig.~\ref{figmixnepall}. \textit{(continued)}.}
%\vspace{-10pt}
\end{figure*}

\clearpage

% Fig 5 (cont), 13 lines
\begin{figure*}[!htb]
%\vspace{-1.5cm}
\begin{center}
\includegraphics[clip=t,width=6.0in]{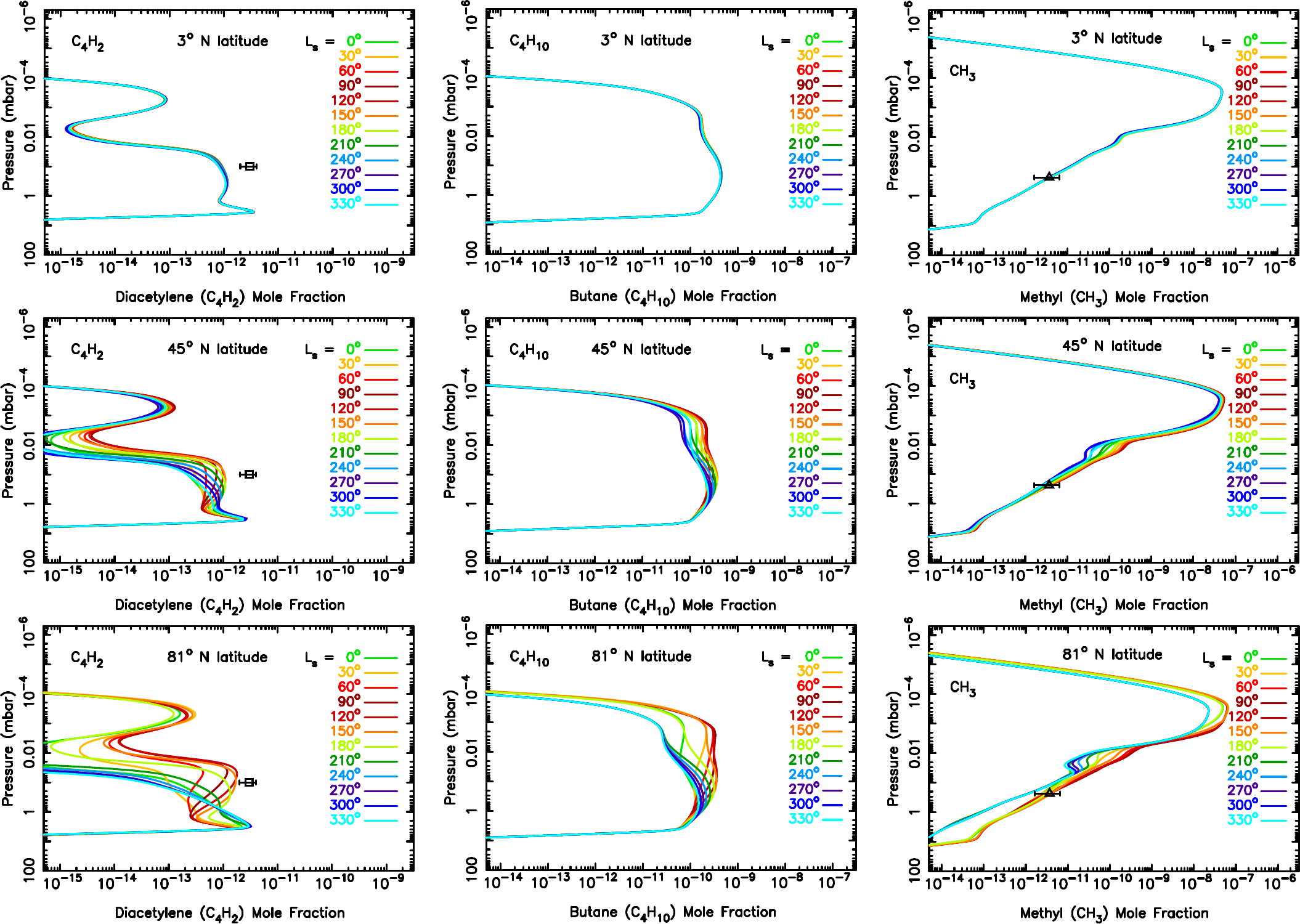}
\end{center}
\vspace{-0.5cm}
\noindent{Fig.~\ref{figmixnepall}. \textit{(continued)}.}
%\vspace{-10pt}
\end{figure*}

\clearpage

\subsection{Seasonal variations in mixing ratios on Neptune}\label{sec:nepseas}

The vertical mixing-ratio profiles of several observed (or potentially observable) hydrocarbons on Neptune are shown 
in Fig.~\ref{figmixnepall} as a function of season for representative low, middle, and high latitudes (top, middle, 
and bottom panels, respectively).  Seasonal variations in the mixing ratios are apparent at altitudes above the 
few-millibar region for most of the photochemical products.  Greater seasonal variation is exhibited at high latitudes 
than low latitudes, simply 
because the seasonal variation in solar insolation is greater at high latitudes than low latitudes (recall Fig.~\ref{figinsol}).  
Note that methane itself does not vary much with either season or latitude because its vertical profile is dominated by eddy 
and molecular diffusion rather than chemistry; the cumulative photochemical products do not rival the abundance of the 
parent CH$_4$ molecules, so photodestruction leads to only a very minor change in the mixing-ratio profile of methane.

% Fig 6, 22 lines
\begin{figure*}[!htb]
%\vspace{-1.5cm}
\begin{center}
\includegraphics[clip=t,scale=0.6]{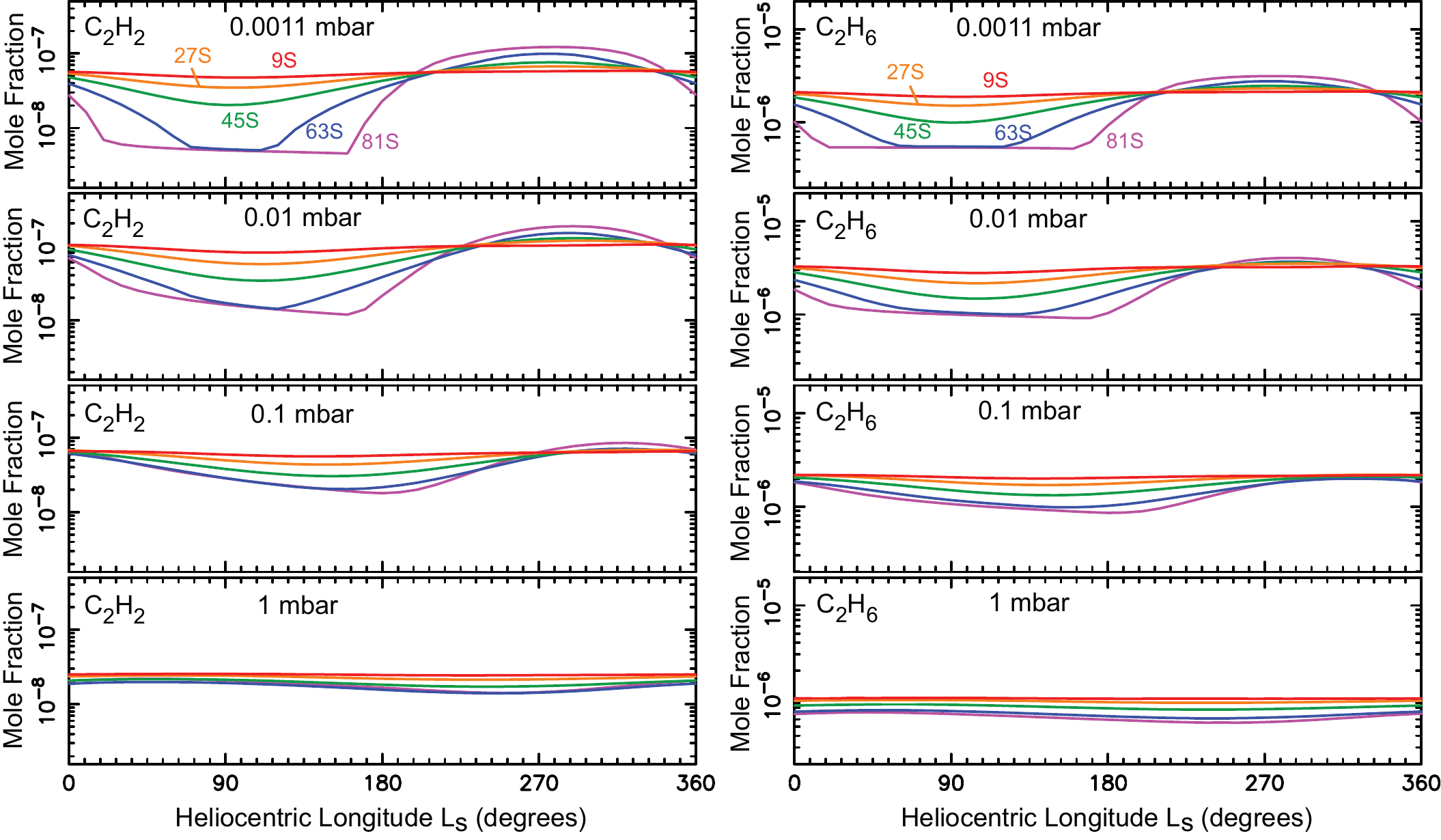}
\end{center}
\vspace{-0.5cm}
\caption{Mixing ratios of acetylene (left) and ethane (right) on Neptune as a function of season (represented by 
solar longitude $L_s$) at 1.1$\scinot-3.$ mbar (top), 0.01 mbar (second from top), 0.1 mbar (third from top), and 
1 mbar (bottom).  Results are shown for five different latitudes: $-9\deg$ (red), $-27\deg$ (orange), $-45\deg$ 
(green), $-63\deg$ (blue), and $-81\deg$ (magenta).  
Seasonal variations are are prominent at lower pressures 
(higher altitudes) but become muted at higher pressures (lower altitudes), and are much greater at high 
latitudes than low latitudes.  The 81\deg\ S latitude region receives no direct sunlight for an extended period during 
the winter, leading to a strong minimum during that period.  Note also that the position of the minimum mixing ratio 
shifts to later in the year as the pressure increases in the atmosphere.
(For interpretation of the references to color in this figure legend, the reader is referred to the online 
version of this article.)
\label{fignepmbar}}
%\vspace{-10pt}
\end{figure*}

\begin{figure*}[!htb]
%\vspace{-1.5cm}
%\begin{tabular}{ll}
%{\includegraphics[clip=t,scale=0.25]{neptimescale_loss}}
%&
%{\includegraphics[clip=t,scale=0.25]{neptimescale_net}}
%$\end{tabular}
\begin{center}
\includegraphics[clip=t,width=5.5in]{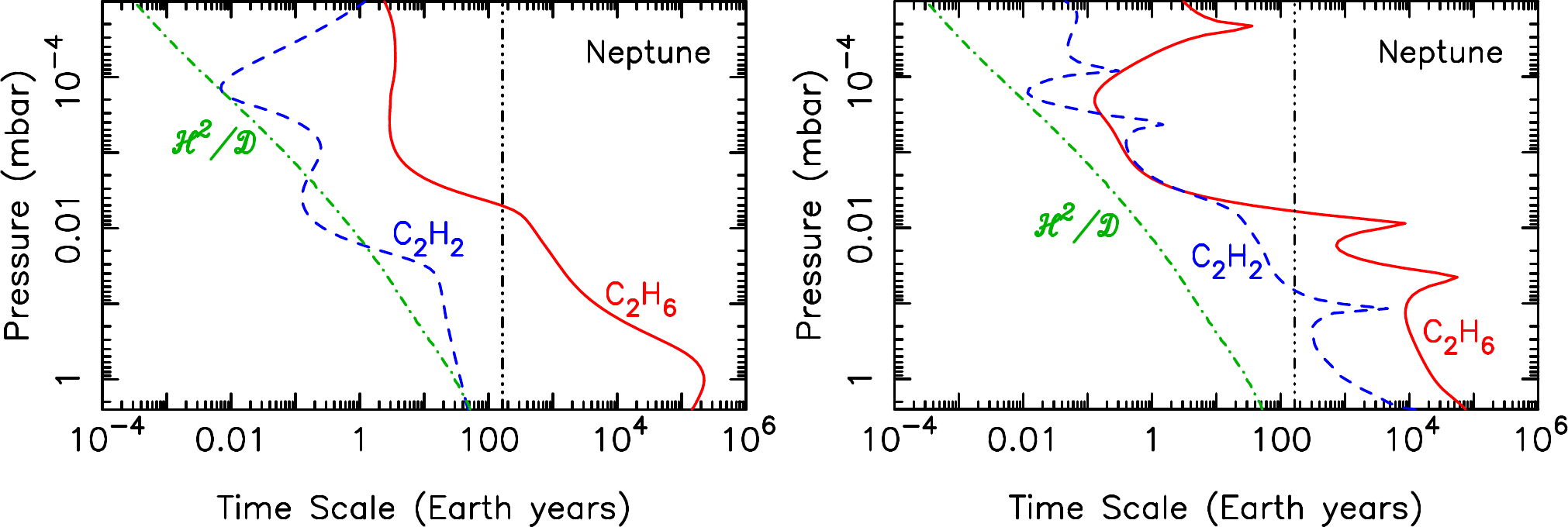}
\end{center}
\vspace{-0.5cm}
\caption{Time constants for photochemical loss (Left) and net photochemical lifetime (Right) above the condensation 
region on Neptune for C$_2$H$_6$ (red solid line) and C$_2$H$_2$ (blue dashed line), along with the diffusion time constant 
for C$_2$H$_6$ (green dot-dashed line, nearly identical to that for C$_2$H$_2$), compared to a Neptune year (vertical 
black triple-dot-dashed line) and a Neptune season (orange dotted line) for $-3\deg$ latitude, vernal equinox.  
The photochemical loss time scale is defined as the species concentration divided by the chemical loss rate; the net 
photochemical time scale is defined as the species concentration divided by the absolute magnitude 
of the chemical production rate minus the loss rate for the species.  The latter gives a better measure of the 
species stability in the presence of efficient recycling processes; the sharp peaks occur where production and loss rates 
are nearly equal.  The diffusion time scale is defined as 
the square of the generalized scale height divided by the generalized diffusion coefficient \citep[see][]{mosesgr05}.  
Note that chemical and diffusion time scales are shorter than a Neptune season in the peak production region 
from 1 to a few $\mu$bar but the net chemical lifetimes become longer than a Neptune season in the middle 
and lower stratosphere.
(For interpretation of the references to color in this figure legend, the reader is referred to the online 
version of this article.)
\label{figneptimescale}}
%\vspace{-10pt}
\end{figure*}

For stable species with relatively long chemical lifetimes, such as C$_2$H$_2$ and C$_2$H$_6$, the seasonal 
variations are confined to high altitudes and become progressively smaller in magnitude at deeper pressures (see also 
Fig.~\ref{fignepmbar}).  This behavior stems from the fact that both the diffusion time scales and chemical time 
constants for these species generally increase with increasing pressure (see Fig.~\ref{figneptimescale}, which
displays the time constants at $-3\deg$ latitude for the vernal equinox situation at the start of the model run).  Chemical 
time constants do vary considerably with both latitude and season, but the general trend of shorter time constants at 
higher altitudes holds true for all situations.  Both the diffusion time scales and chemical lifetimes are less than a Neptune
season in the region from $\sim$1 to a few $\mu$bar, where the C$_2$H$_2$ and C$_2$H$_6$ production rates peak in the 
upper stratosphere.  Therefore, at $\sim$0.001 mbar, the C$_2$H$_2$ and C$_2$H$_6$ abundances respond quickly 
to changes in insolation.  The increased insolation during the summer season leads to increased C$_2$H$_2$ and C$_2$H$_6$ 
production rates and an increased abundance of hydrocarbon photochemical products in general (see Figs.~\ref{figmixnepall} \& 
\ref{fignepmbar}).  The short vertical diffusion time scales ensure that the photochemically produced species are 
transported rapidly to lower altitudes.  Conversely, lower photochemical production rates in the winter lead to lower 
abundances of hydrocarbons at high altitudes, as the molecules produced in previous seasons have already been transported 
away.  Photolysis by direct solar ultraviolet photons shuts off during the long polar winter, causing a strong dip in the 
high-altitude C$_2$H$_2$ and C$_2$H$_6$ abundances at high latitudes in the winter season (see Figs.~\ref{figmixnepall} 
\& \ref{fignepmbar}), which quickly recovers as soon as the region receives direct sunlight again.  However, solar 
Lyman $\alpha$ radiation scattered by H atoms in the local interplanetary medium continues to contribute to CH$_4$ 
photolysis even in the polar night, preventing a much stronger depletion in hydrocarbon abundances.  

% Fig 7, 27 lines

At 0.01 mbar, the chemical lifetime of C$_2$H$_2$ approaches a Neptune season and that of C$_2$H$_6$ exceeds 
a Neptune year (Fig.~\ref{figneptimescale}).  The C$_2$H$_2$ and C$_2$H$_6$ mixing ratios in this pressure region 
are strongly influenced by molecules being transported into the region from higher altitudes, albeit with a 
time delay, which introduces a phase lag into the seasonal behavior.  Seasonal variations are still apparent, 
but become more muted, and the minimum and maximum in the mixing ratios are shifted away from the solstices toward 
later times in the winter and summer seasons.  Those phase lags continue to grow with depth, and the seasonal 
variations become reduced in magnitude, until almost no seasonal variations remain apparent in the few mbar 
region.  At that point, the yearly averaged solar insolation controls the overall abundances, rather than the 
seasonally variable insolation.  Other relatively chemically stable, long-lived species such as C$_3$H$_{8}$ 
and C$_4$H$_{10}$ exhibit similar behavior as C$_2$H$_2$ and C$_2$H$_6$, with seasonal variations confined to 
higher altitudes, and phase lags introduced in the $\sim$0.01-10 mbar region.

Less stable species such as C$_2$H$_4$, CH$_3$C$_2$H, C$_4$H$_2$, and CH$_3$ that have shorter chemical lifetimes 
at depth continue to experience seasonal variations at all altitudes above their condensation regions (i.e., 
for the first three aforementioned species, which condense).  The seasonal behavior of these species can be 
complicated, as both diffusion and 
\textit{in situ} photochemistry affect the behavior, photodestruction of other seasonally variable species such as 
C$_2$H$_2$ and C$_2$H$_6$ contribute to their production, and photochemically produced radicals such as atomic H 
affect both production and destruction \citep[see][for details]{moses00a,moses05,moses15}.  This more complicated 
chemistry (as well as uncertainties in the thermal structure for the observational abundance determinations) 
contributes to the apparent disagreements between the CH$_3$C$_2$H and C$_4$H$_2$ abundances predicted by the 
models and inferred from observations.  The model-data mismatch for these species is not of great concern 
here, because each chemical rate coefficient in the model is subject to uncertainties, which when combined, lead to 
factors of a few uncertainties in the predicted abundances \citep[e.g.,][]{dobrijevic03,dobrijevic10}.  Accounting for 
these uncertainties in our seasonal models is beyond the scope of this paper.

Full results 
for the abundances of all species in the model as a function of latitude and season can be found in the 
Supplementary Material for this article.

\subsection{Seasonal variations in mixing ratios on Uranus}

\begin{figure*}[!htb]
%\vspace{-1.5cm}
\begin{center}
\includegraphics[clip=t,width=6.0in]{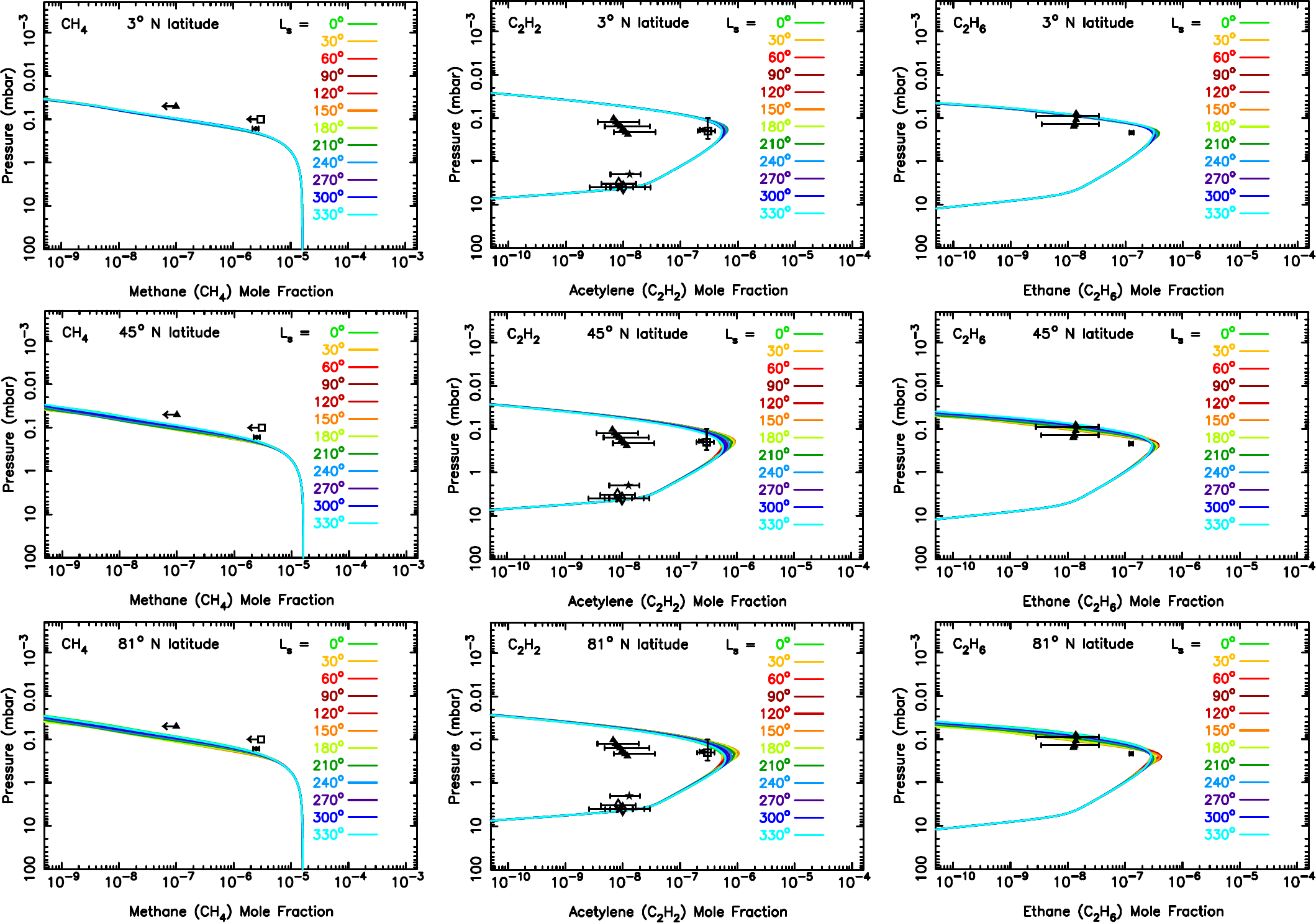}
\end{center}
\vspace{-0.5cm}
\caption{Mixing ratio profiles for key hydrocarbons (as labeled) on Uranus as 
a function of season for 3\deg N planetocentric latitude (Top), 45\deg\ planetocentric latitude (Middle), 
and 81\deg\ planetocentric latitude (Bottom).  Seasons are shown every 30\deg\ in $L_s$.
Data points are from various published observations
\citep{encrenaz86,encrenaz98,orton87,herbert87,caldwell88,yelle89,bishop90,bezard99,orton14chem}.
\label{figmixuranall}}
%\vspace{-10pt}
\end{figure*}

\begin{figure*}[!htb]
%\vspace{-1.5cm}
\begin{center}
\includegraphics[clip=t,scale=0.6]{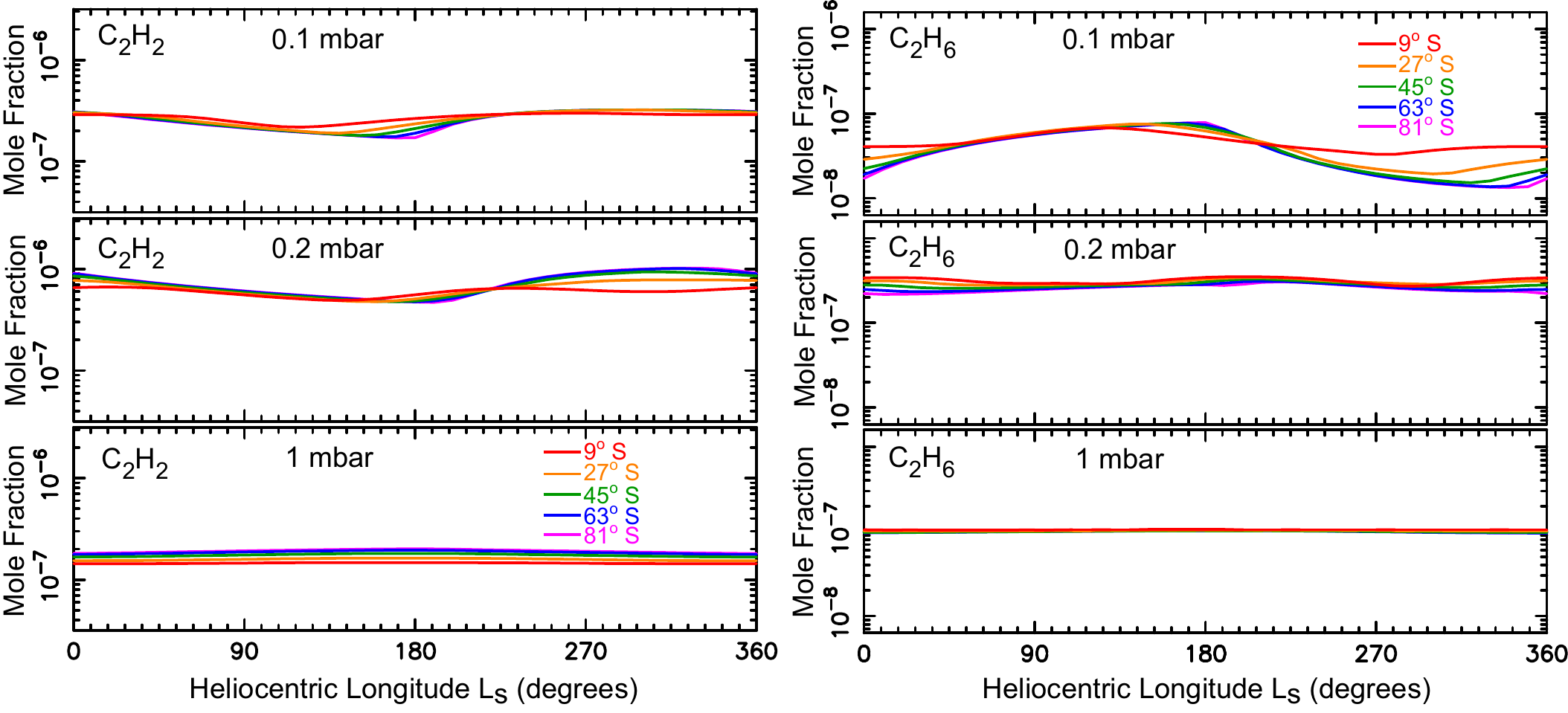}
\end{center}
\vspace{-0.5cm}
\caption{Mixing ratios of acetylene (left) and ethane (right) on Uranus as a function of season (represented by 
solar longitude $L_s$) at 0.1 mbar (top), 0.2 mbar (middle), and 1 mbar (bottom).  Results are shown for five 
different latitudes: $-9\deg$ (red), $-27\deg$ (orange), $-45\deg$ (green), $-63\deg$ (blue), and $-81\deg$ (magenta).
Seasonal variations are muted in general on Uranus but are more pronounced at lower pressures (higher altitudes).
(For interpretation of the references to color in this figure legend, the reader is referred to the online 
version of this article.)
\label{figuranmbar}}
%\vspace{-10pt}
\end{figure*}

\begin{figure*}[!htb]
%\vspace{-1.5cm}
%\begin{tabular}{ll}
%{\includegraphics[clip=t,scale=0.25]{urantimescale_loss}}
%&
%{\includegraphics[clip=t,scale=0.25]{urantimescale_net}}
%\end{tabular}
%\vspace{-0.5cm}
\begin{center}
\includegraphics[clip=t,width=5.5in]{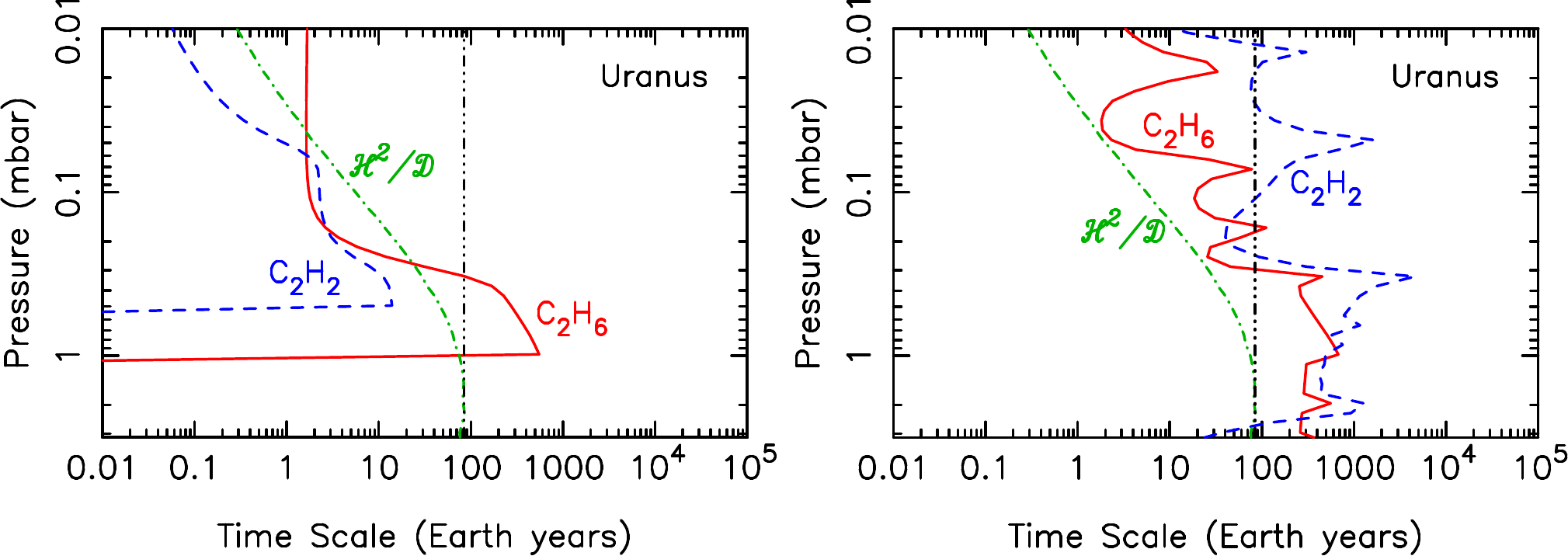}
\end{center}
\vspace{-0.5cm}
\caption{Time constants for photochemical loss (Left) and net photochemical lifetime (Right) above the condensation 
region on Uranus for C$_2$H$_6$ (red solid line) and C$_2$H$_2$ (blue dashed line), along with the diffusion time constant 
for C$_2$H$_6$ (green dot-dashed line, nearly identical to that for C$_2$H$_2$), compared to a Uranus year (vertical 
black triple-dot-dashed line) and a Uranus season (orange dotted line) for $-3\deg$ latitude, vernal equinox.  
%The photochemical loss time scale is defined as the species concentration divided by the chemical loss rate; the net 
%photochemical time scale is defined as the species concentration divided by the absolute magnitude 
%of the chemical production rate minus the loss rate for the species.  The latter gives a better measure of the 
%species stability given efficient recycling processes.  The diffusion time scale is defined as 
%the square of the generalized scale height divided by the generalized diffusion coefficient \citep[see][]{mosesgr05}.  
%Note that chemical and diffusion time scales are shorter than a Neptune season in the peak production region 
%from 1 to a few $\mu$bar but the chemical time constants become longer than a Neptune season in the middle 
%and lower stratosphere.
(For interpretation of the references to color in this figure legend, the reader is referred to the online 
version of this article.)
\label{figurantimescale}}
%\vspace{-10pt}
\end{figure*}

The mixing-ratio profiles for several observed (or potentially observable) hydrocarbons on Uranus are shown in 
Fig.~\ref{figmixuranall} as a function of season for representative low, middle, and high latitudes.  The first 
thing to note from Fig.~\ref{figmixuranall} is that seasonal variations in hydrocarbon abundances are minor on 
Uranus compared to Neptune.  Small changes in the C$_2$H$_2$ and C$_2$H$_6$ mixing ratio with season are 
predicted in the $\sim$0.1--1 mbar region (see also Fig.~\ref{figuranmbar}), with virtually no seasonal change 
at deeper pressures.  Vertical transport on Uranus is not very vigorous.  The methane homopause is therefore 
located at a relatively deep $\sim$7$\scinot-2.$ mbar on Uranus --- a pressure roughly three orders of magnitude 
greater than the homopause level on Neptune --- and so the CH$_4$ is not carried as high up in the stratosphere 
on Uranus as it is on Neptune.  Both C$_2$H$_2$ and C$_2$H$_6$ have their peak production region near $\sim$0.2 
mbar on Uranus, at which point chemical time constants exceed (and diffusion time constants approach) a Uranus season (see 
Fig.~\ref{figurantimescale}).  At altitudes above the 0.2-mbar level, the C$_2$H$_2$ production rate exceeds its loss rate, 
and because more C$_2$H$_2$ is produced in the summer season and less in the winter, the C$_2$H$_2$ mixing ratio 
at 0.1 mbar is greatest in the summer-to-fall time frame and least in the winter-to-spring, with the phase lag due to 
the long diffusion and chemistry time constants being apparent in Fig.~\ref{figuranmbar}.  In contrast, the loss rate of 
C$_2$H$_6$ from photolysis exceeds the production rate at altitudes above the 0.2-mbar level, so more of the C$_2$H$_6$ 
that is being carried upward from its peak production region is destroyed in the summer than the winter, and 
the mixing ratio of C$_2$H$_6$ is greatest in the winter/spring and smallest in the summer/fall.  At pressures of 
1 mbar and greater, diffusion and chemical time constants are so long that no seasonal changes are expected for C$_2$H$_2$ 
and C$_2$H$_6$ at these pressures.

% Fig 8, 18 lines

\clearpage

% Fig 8 (cont), 12 lines
\begin{figure*}[!htb]
%\vspace{-1.5cm}
\begin{center}
\includegraphics[clip=t,width=6.0in]{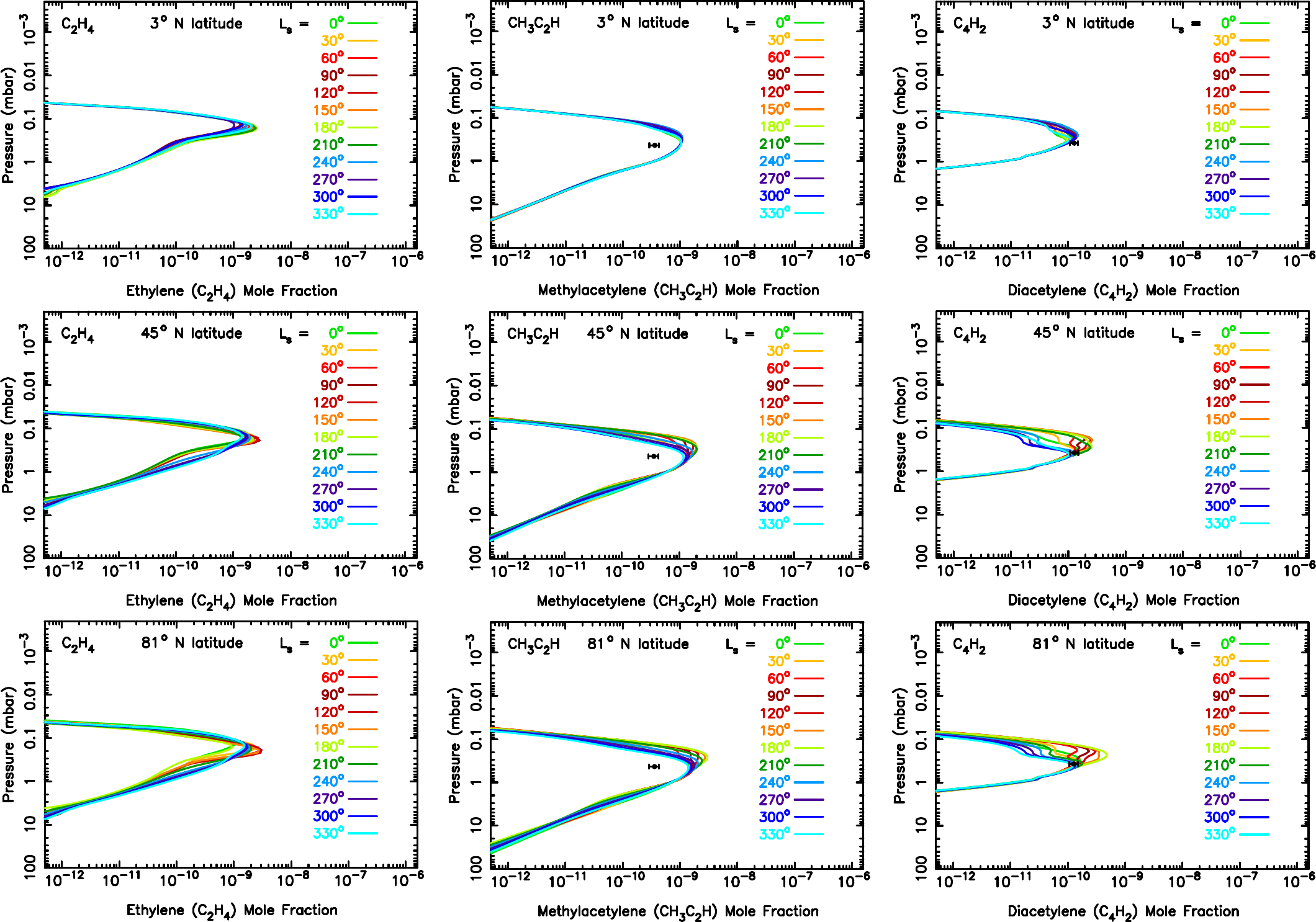}
\end{center}
\vspace{-0.5cm}
\noindent{Fig.~\ref{figmixuranall}. \textit{(continued)}.}
%\vspace{-10pt}
\end{figure*}

% Fig 9, 17 lines

Even the species with shorter chemical lifetimes, such as C$_2$H$_4$ and CH$_3$C$_2$H experience 
relatively little seasonal variation in comparison with Neptune, because of the long diffusion time scales and lack 
of presence of these molecules at high altitudes where time scales are typically shorter.  We would therefore not 
expect to observe much in the way of seasonal variations in hydrocarbon abundances on Uranus unless vertical 
transport --- or stratospheric circulation in general --- changes with season.  One exception is C$_4$H$_2$, whose 
mixing ratio varies by as much as an order of magnitude in the $\sim$0.2--0.5 mbar region (see Fig.~\ref{figmixuranall}).  

Note, however, that if stratospheric temperatures vary with season, the column abundance of all the condensable 
hydrocarbons will be greater at latitudes and seasons where the lower-stratospheric temperatures are warmer, because 
the available region over which the molecules can remain in the vapor phase expands to deeper pressures 
(i.e., the condensation region becomes more narrowly confined to the coldest pressure levels surrounding the 
tropopause).  On Uranus, where the stratosphere is colder and the column density of hydrocarbons is so low anyway 
due to sluggish atmospheric mixing, changes in the pressure level of condensation due to seasonal temperature 
changes \citep[e.g.,][]{conrath90} may have more of an effect on observed hydrocarbon abundances than changes 
due to seasonal photochemistry itself.  This effect applies to all hydrocarbons that can condense in the lower 
stratosphere, including C$_2$H$_2$ and C$_2$H$_6$, and would be observable if remote-sensing observations probe 
deep enough to sense the condensation region near the $\sim$1--10 mbar level.

% Fig 10, 27 lines

%\clearpage

\subsection{Meridional variations in mixing ratios on Uranus and Neptune}\label{sec:merid}

\begin{figure*}[!htb]
%\vspace{-1.5cm}
%\begin{tabular}{ll}
%{\includegraphics[clip=t,scale=0.32]{nepc2h2c2h6vslat}}
%&
%{\includegraphics[clip=t,scale=0.32]{nepc3h4c4h2vslat}}
%\end{tabular}
%\vspace{-0.5cm}
\begin{center}
\includegraphics[clip=t,width=5.5in]{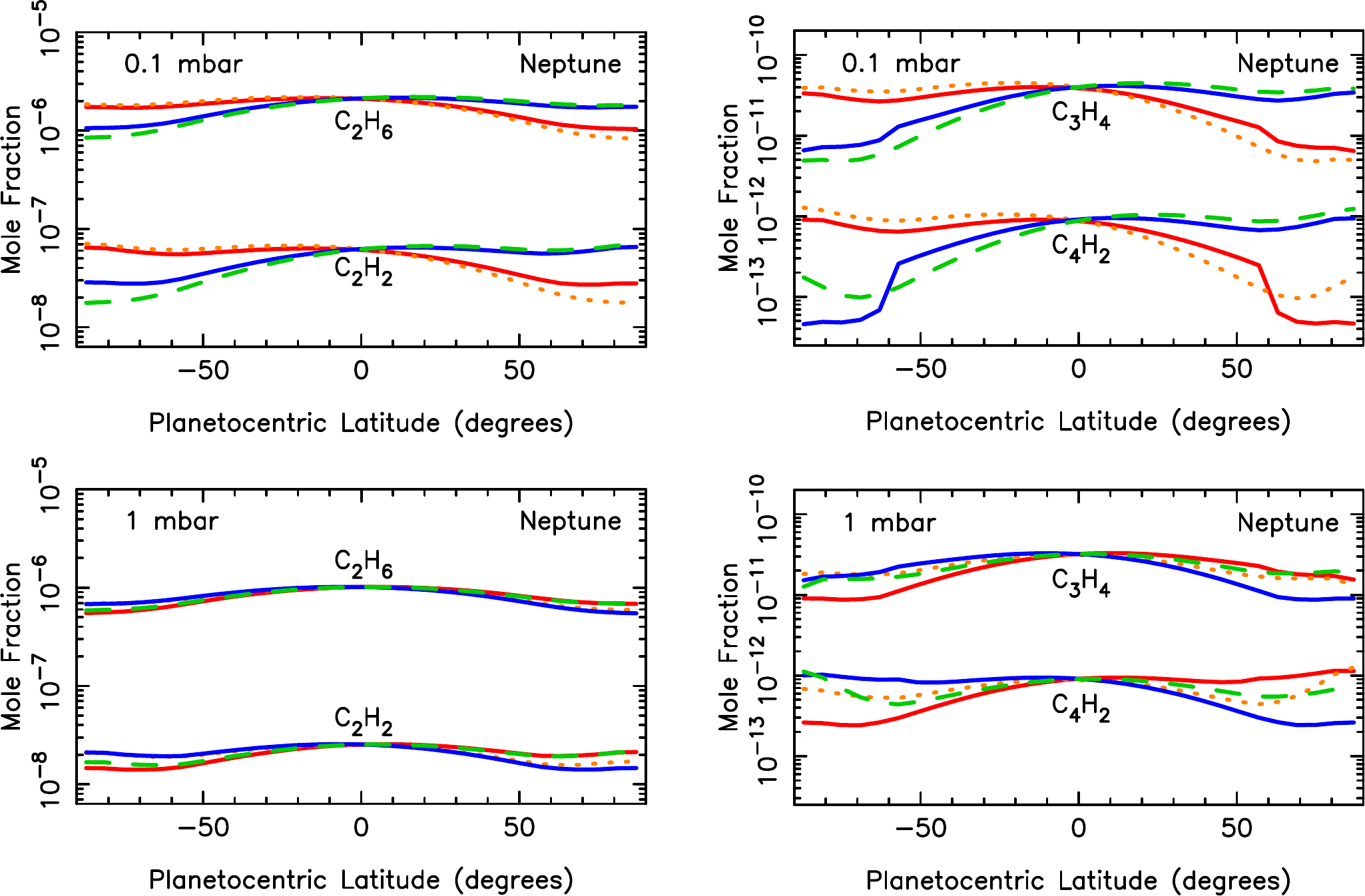}
\end{center}
\vspace{-0.5cm}
\caption{Mixing ratios of ethane and acetylene (Left) and methylacetylene and diacetylene (Right) at 0.1 mbar (Top) and 
1 mbar (bottom) on Neptune for four different seasons: $L_s$ = 0\deg\ (dotted orange), 90\deg\ (solid blue), 180\deg\ 
(dashed green), and 270\deg\ (solid red).
(For interpretation of the references to color in this figure legend, the reader is referred to the online 
version of this article.)
\label{fignepvslat}}
%\vspace{-10pt}
\end{figure*}

As with seasonal variations, latitude variations in mixing ratios are more pronounced at higher altitudes.  
Fig.~\ref{fignepvslat} illustrates how the C$_2$H$_2$, C$_2$H$_6$, CH$_3$C$_2$H, and C$_4$H$_2$ mixing ratios 
on Neptune vary as a function of latitude at four specific seasons: $L_s$ = 0\deg\ (dotted orange curves), 
90\deg\ (solid blue curves), 180\deg\ (dashed green curves), and 270\deg\ (solid red curves).  At high altitudes 
(e.g., pressures less than $\sim$0.1 mbar), hemispheric dichotomies are readily apparent, typically with fewer 
photochemical products being present in the hemisphere experiencing winter and spring (due to the long 
accumulated time periods with low sunlight), and more photochemical products in the hemisphere experiencing 
summer and fall.  At higher pressures, the mixing ratios of longer-lived species such as C$_2$H$_2$ and 
C$_2$H$_6$ become more symmetric about the equator, with the annual average actinic flux dominating the 
latitude variations (e.g., recall Fig.~\ref{figannual}).  On Neptune, the variations in annual average solar 
irradiation with latitude lead to a greater 1-mbar abundance of C$_2$H$_2$, C$_2$H$_6$, and CH$_3$C$_2$H at low 
latitudes in comparison with high latitudes,  while species with short chemical lifetimes, such as C$_4$H$_2$, 
continue to exhibit hemispheric asymmetries at pressures of 1 mbar (see Fig.~\ref{fignepvslat}).  Note also for 
the shorter-lived species that the increasing phase lag with depth can lead to hemispheric asymmetries at 1 mbar 
that are completely out of phase with those at 0.1 mbar.  This effect is important to keep in mind when considering 
observations that are more sensitive to deeper levels --- the abundances are out of phase at these depths in 
comparison to the instantaneous seasonal forcing, and the vertical profiles change significantly with season and 
latitude, which is important to keep in mind when choosing priors for retrievals.

% Fig 11, 18 lines

\begin{figure*}[!htb]
%\vspace{-1.5cm}
%\begin{tabular}{ll}
%{\includegraphics[clip=t,scale=0.32]{uranc2h2c2h6vslat}}
%&
%{\includegraphics[clip=t,scale=0.32]{uranc3h4c4h2vslat}}
%\end{tabular}
\begin{center}
\includegraphics[clip=t,width=5.5in]{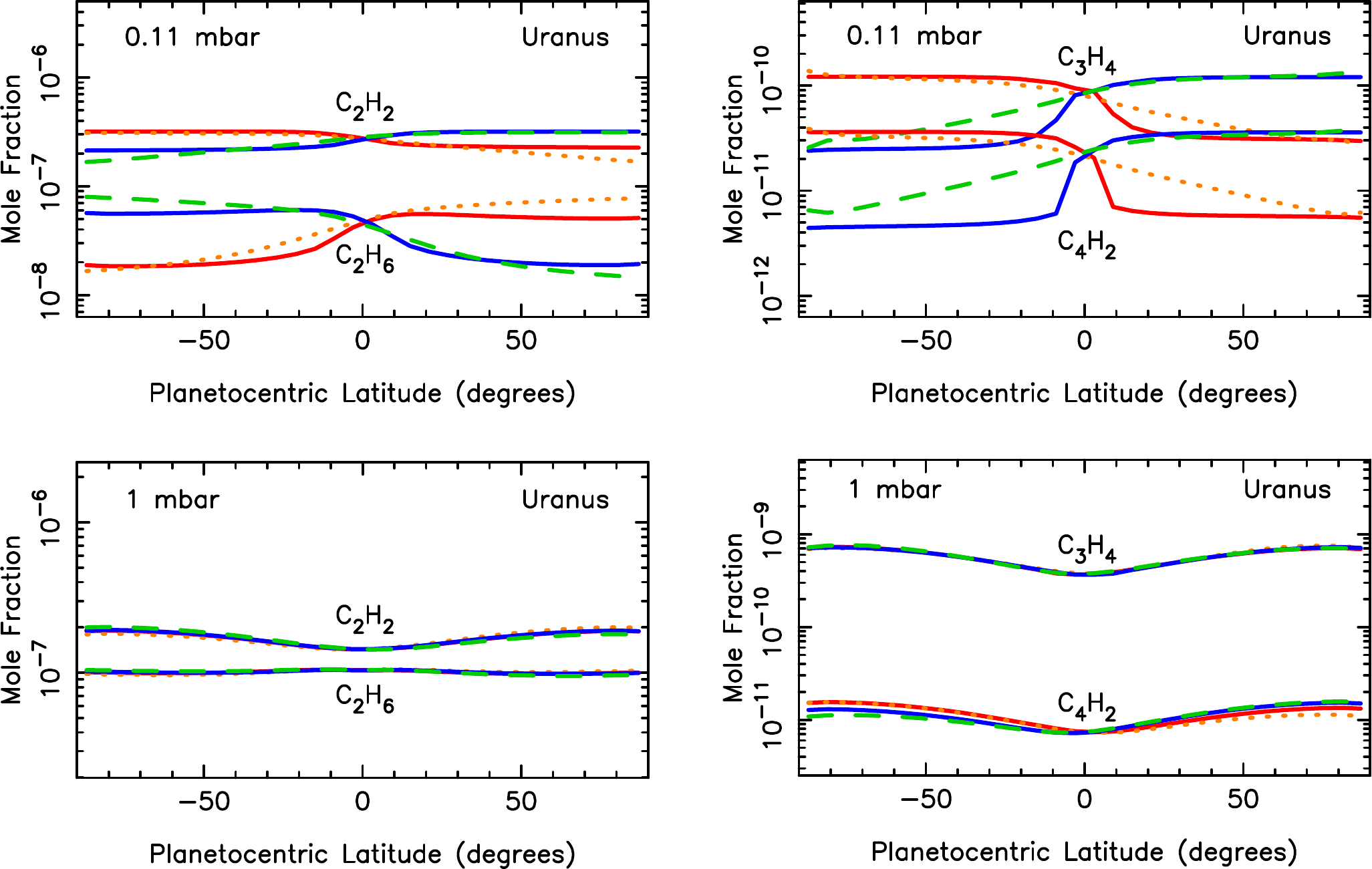}
\end{center}
\vspace{-0.5cm}
\caption{Same as Fig.~\ref{fignepvslat}, except for Uranus.
\label{figuranvslat}}
%\vspace{-10pt}
\end{figure*}

The sluggish vertical mixing and large axial tilt on Uranus result in notable differences in latitude 
variations on Uranus (see Fig.~\ref{figuranvslat}), in comparison to Neptune.  The longer vertical diffusion time 
scales and shorter year on Uranus result in seasonal variations being damped at 1 mbar on Uranus, and even species 
with relatively short photochemical lifetimes, like C$_3$H$_4$ and C$_4$H$_2$, are controlled by the annual 
average actinic flux at these pressures.  Because the annual average daily insolation is greater at the poles 
than the equator on the highly tilted Uranus (Fig.~\ref{figannual}), most photochemically produced species at 
1 mbar have a larger mixing ratio at high latitudes than low latitudes.  Ethane is an exception.  The low-altitude 
homopause on Uranus results in different dominant hydrocarbon reactions in the region in which methane is photolyzed 
than on the other giant planets.  Photolysis reactions control both the production and loss of ethane in the 
homopause region, which act to counteract each other, and there are no effective in situ chemical loss processes 
for C$_2$H$_6$ at 1 mbar.  The net column production rate for C$_2$H$_6$ therefore does not vary much as a function 
of latitude.  The C$_2$H$_6$ that is produced near the homopause region diffuses downward, and because we have 
assumed that the eddy diffusion coefficient profile does not vary with latitude, the ethane abundance at depth 
on Uranus is relatively constant with latitude.

% Fig 12, 14 lines

At higher altitudes on Uranus (e.g., 0.1 mbar in Fig.~\ref{figuranvslat}), the photochemically produced 
species exhibit hemispheric dichotomies, with most species being more abundant in the summer-to-fall 
hemisphere than the winter-to-spring hemisphere, as on Neptune.  However, C$_2$H$_6$ is again an exception.  
At the CH$_4$ homopause on Uranus, CH$_4$ is less abundant than on the other giant planets and is 
less effective at shielding the C$_2$H$_6$ from photolysis.  Seasonal variations in solar irradiation 
therefore lead to a reduction in the high-altitude C$_2$H$_6$ abundance in the summer/fall hemisphere 
in comparison with the winter/spring hemisphere.

Keep in mind that the models presented in this paper do not consider meridional circulation or differences 
in vertical transport or methane abundance with latitude.  Such processes could end up controlling latitude 
variations in photochemically produced species abundances on the giant planets, particularly on Uranus, 
where chemical and diffusion time constants are long and latitudinal/seasonal variations are predicted 
to be weak.

\subsection{Variations in column abundance with latitude and season on Neptune}\label{sec:columnnep}

\begin{figure*}[!htb]
%\vspace{-1.5cm}
\begin{center}
\includegraphics[clip=t,width=4.0in]{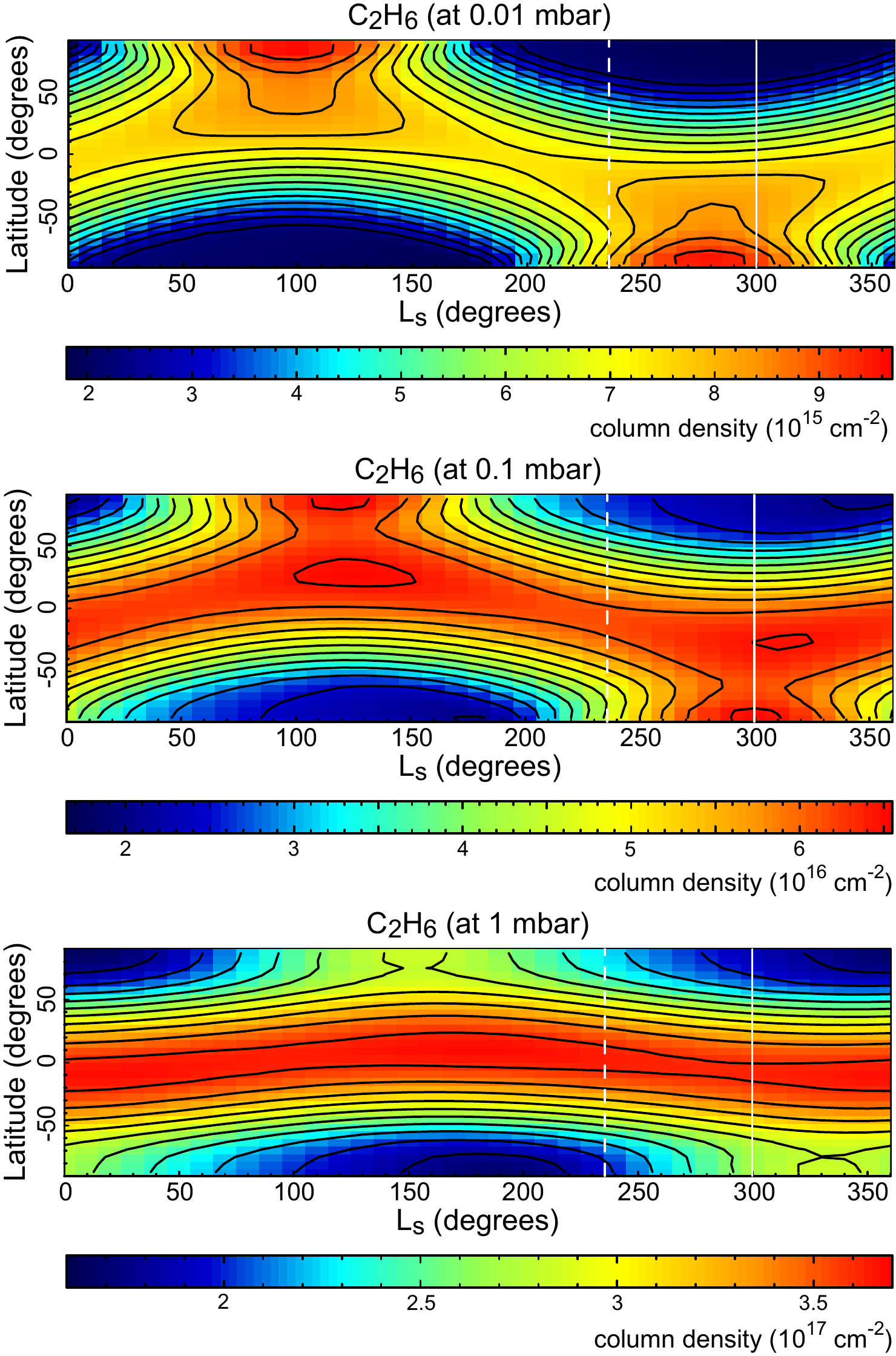}
\end{center}
\vspace{-0.5cm}
\caption{The column abundance of ethane on Neptune above 0.01 mbar (Top), 0.1 mbar (Middle), and 
1 mbar (Bottom), as a function of planetocentric latitude and season ($L_s$).  The dashed vertical 
line represents the time of the \textit{Voyager} encounter, and the solid vertical line represents 
7 September 2018, the date of the next Neptune opposition.
\label{fignepcontc2h6}}
%\vspace{-10pt}
\end{figure*}

\begin{figure*}[!htb]
%\vspace{-1.5cm}
\begin{center}
\includegraphics[clip=t,width=4.0in]{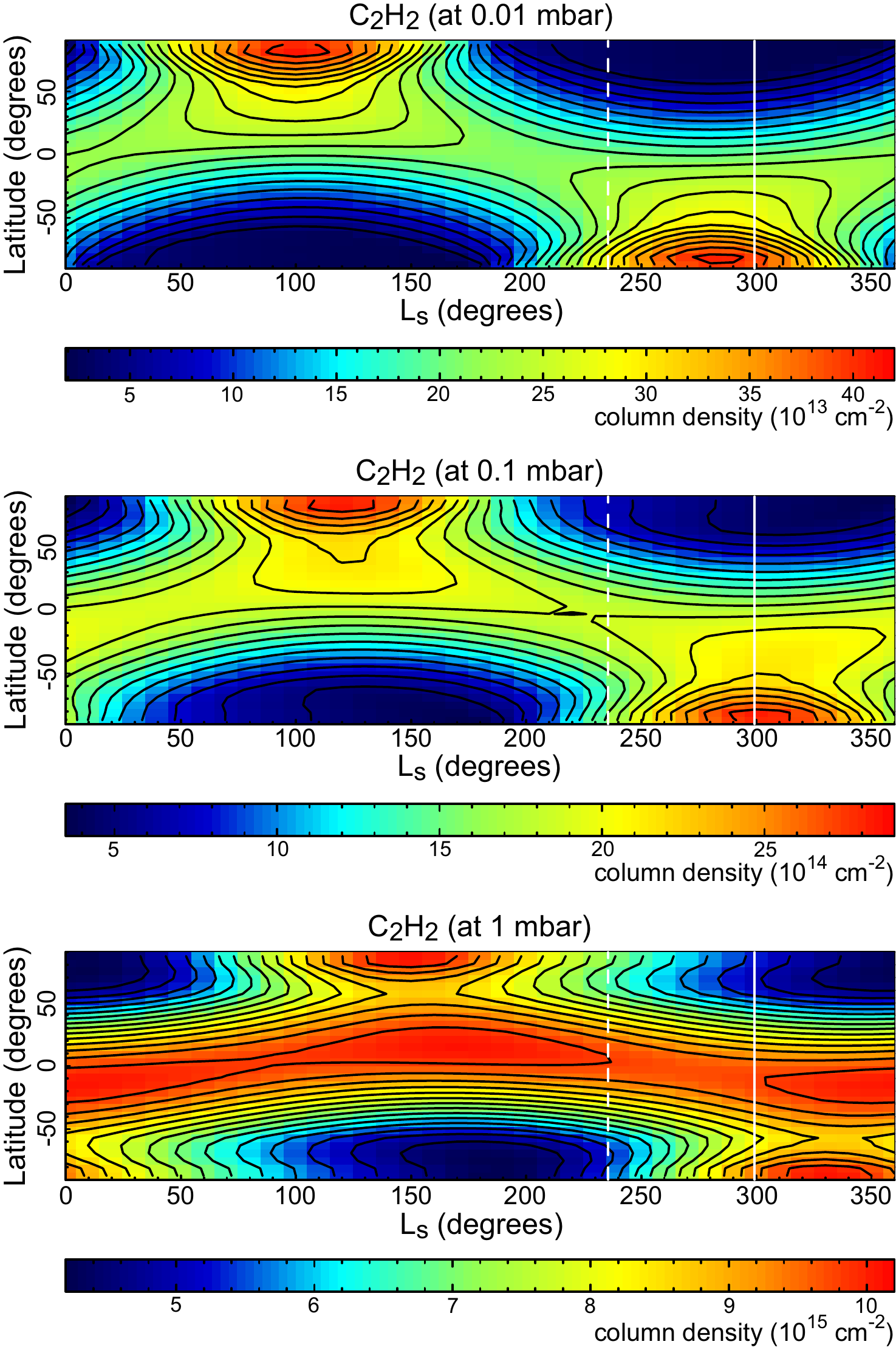}
\end{center}
\vspace{-0.5cm}
\caption{Same as Fig.~\ref{fignepcontc2h6}, except for acetylene.
\label{fignepcontc2h2}}
%\vspace{-10pt}
\end{figure*}

\begin{figure*}[!htb]
%\vspace{-1.5cm}
\begin{center}
\includegraphics[clip=t,width=4.0in]{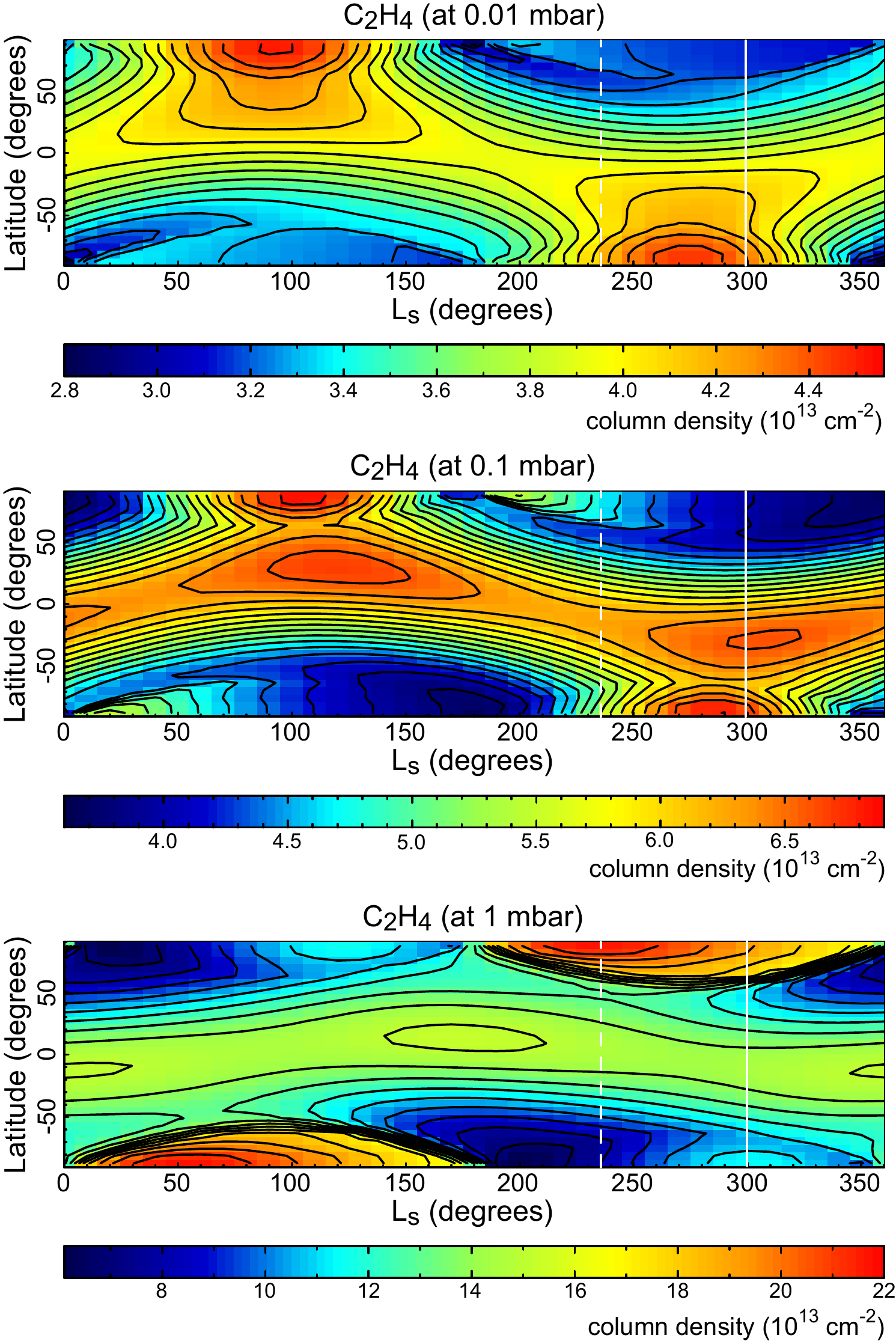}
\end{center}
\vspace{-0.5cm}
\caption{Same as Fig.~\ref{fignepcontc2h6}, except for ethylene.
\label{fignepcontc2h4}}
%\vspace{-10pt}
\end{figure*}

\begin{figure*}[!htb]
%\vspace{-1.5cm}
\begin{center}
\includegraphics[clip=t,width=4.0in]{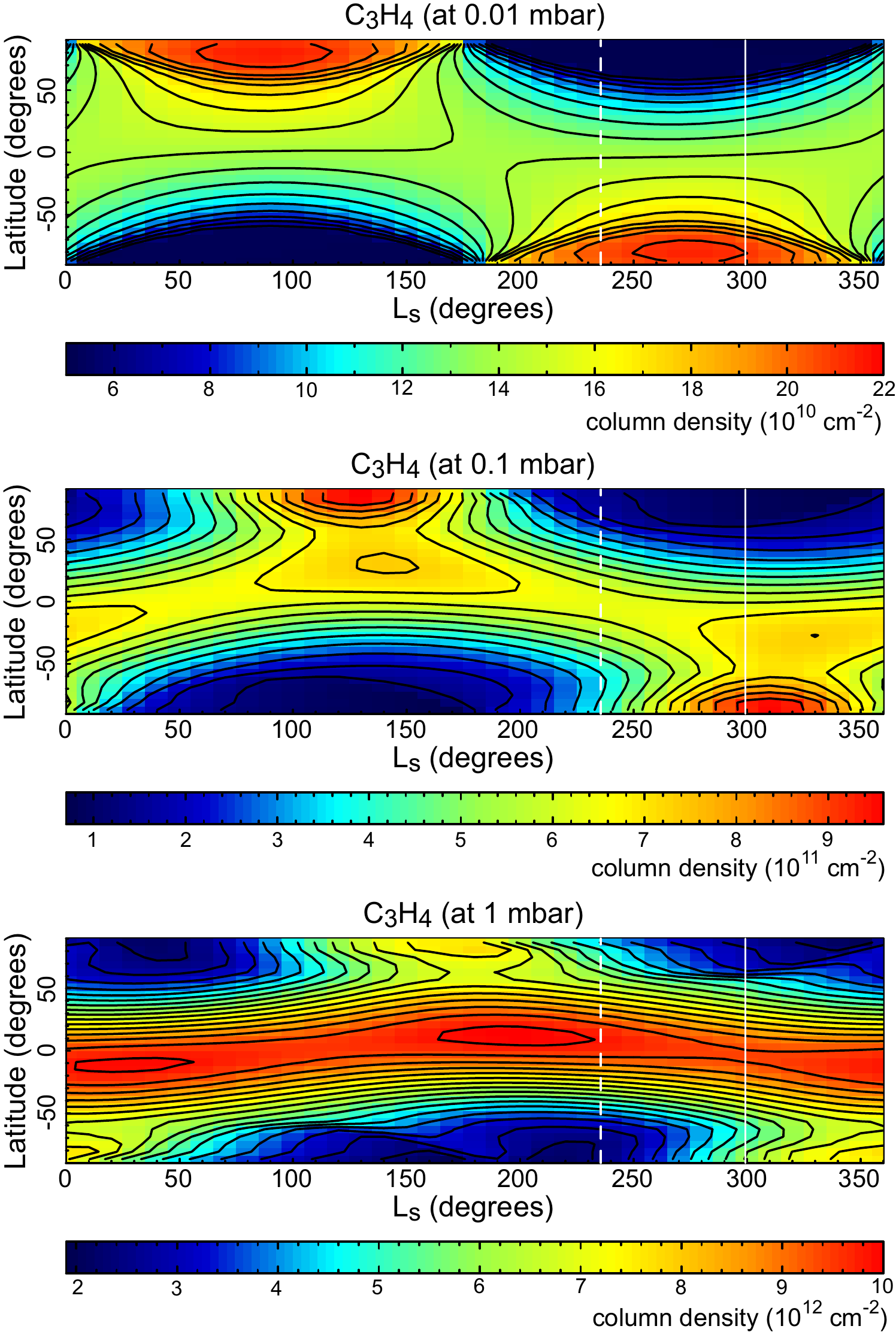}
\end{center}
\vspace{-0.5cm}
\caption{Same as Fig.~\ref{fignepcontc2h6}, except for methylacetylene.
\label{fignepcontc3h4}}
%\vspace{-10pt}
\end{figure*}

\begin{figure*}[!htb]
%\vspace{-1.5cm}
\begin{center}
\includegraphics[clip=t,width=4.0in]{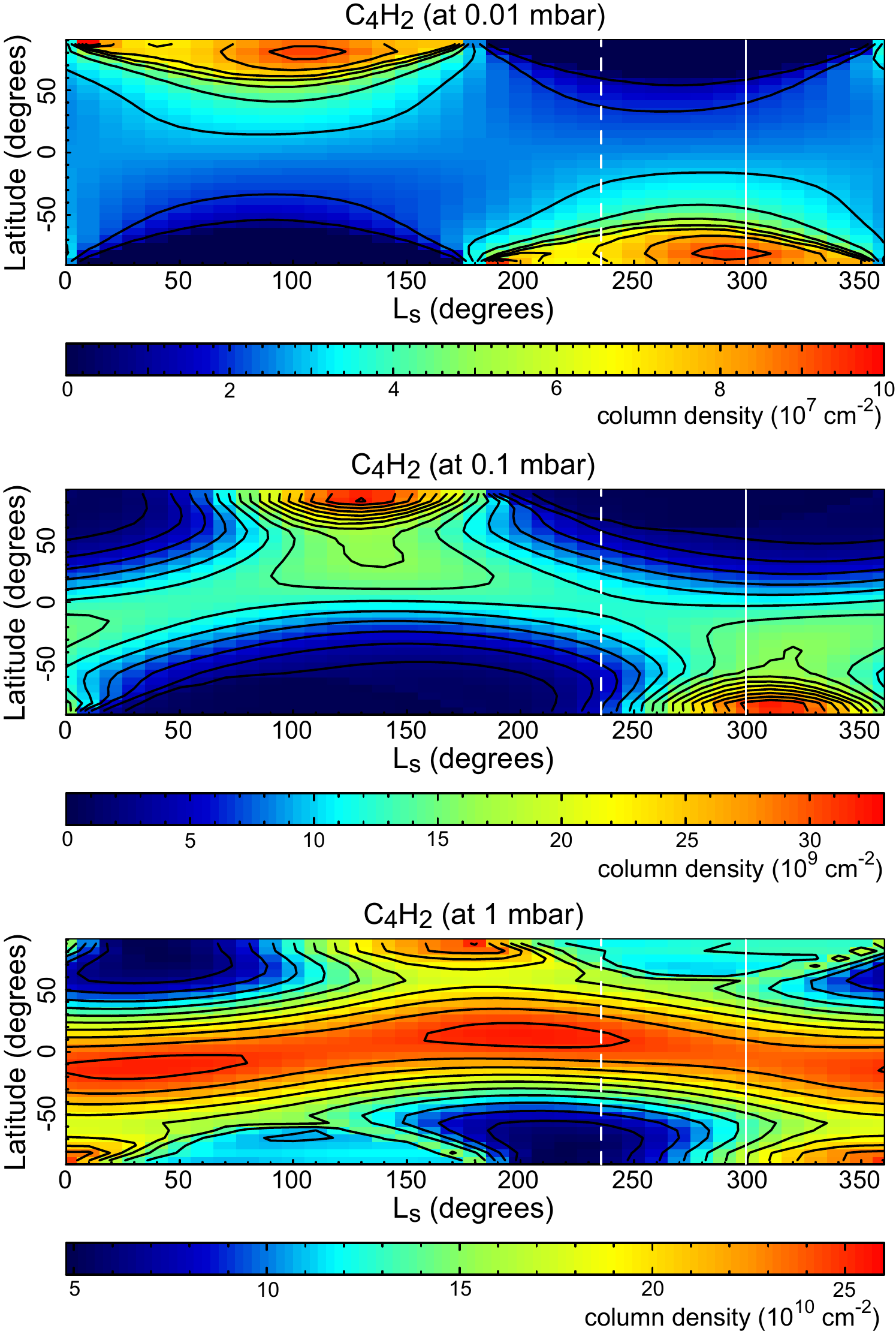}
\end{center}
\vspace{-0.5cm}
\caption{Same as Fig.~\ref{fignepcontc2h6}, except for diacetylene.
\label{fignepcontc4h2}}
%\vspace{-10pt}
\end{figure*}

Because the hydrocarbon mixing ratios change significantly with altitude and the because the observed emission is 
proportional to the column abundance of the molecules when the emission lines are optically thin, it is convenient 
to consider the column abundance above various pressure regions when predicting observable changes as a function 
of latitude and time.
Figs.~\ref{fignepcontc2h6}--\ref{fignepcontc4h2} show how the column abundances of various hydrocarbons above 
different pressure levels vary as a function of latitude and season on Neptune.

% Fig 13, 17 lines

\clearpage

% Fig 14, 14 lines

\clearpage

% Fig 15, 14 lines

\clearpage

% Fig 16, 14 lines

\clearpage

% Fig 17, 14 lines

\clearpage

As was discussed in section~\ref{sec:nepseas}, the abundances of most of the hydrocarbon 
photochemical products closely track the seasonal variation of solar insolation at high altitudes, 
causing the hydrocarbon column densities above 0.01 mbar to be much greater in the summer hemisphere 
in comparison to the winter hemisphere.  However, chemical and diffusion time scales increase 
with depth in the atmosphere, leading to increasing phase lags in the abundance maxima and minima 
with depth at any particular latitude, along with meridional profiles of column abundance that become more 
symmetric about the equator at greater depths.  These changes are readily apparent for C$_2$H$_6$ 
(Fig.~\ref{fignepcontc2h6}), C$_2$H$_2$ (Fig.~\ref{fignepcontc2h2}), and CH$_3$C$_2$H 
(Fig.~\ref{fignepcontc3h4}).  

Similar trends are seen for C$_2$H$_4$ (Fig.~\ref{fignepcontc2h4}) and C$_4$H$_2$ (Fig.~\ref{fignepcontc4h2}), 
although these two species also exhibit some additional interesting behavior.  Note from Fig.~\ref{fignepcontc2h4} 
that the C$_2$H$_4$ column abundance above 1 mbar begins to increase strongly during the high-latitude winter months.  
This behavior results from the fact that photolysis is normally an effective loss process for C$_2$H$_4$ in the 
0.1-10 mbar region, but direct solar UV photolysis is absent during the winter polar night.  Although the stellar 
background UV source and scattered LIPM Lyman alpha photons are still available at that time, the overall 
C$_2$H$_4$ photolysis rate is greatly diminished, and the local production rate of C$_2$H$_4$ exceeds its loss rate.
The dominant production of C$_2$H$_4$ at this time occurs through C$_2$H$_2$ + H + M $\rightarrow$ C$_2$H$_3$ + M, followed by 
C$_2$H$_3$ + H$_2$ $\rightarrow$ C$_2$H$_4$ + H, causing a net conversion of C$_2$H$_2$ into C$_2$H$_4$ during the 
long winter polar nights at $\sim$0.1-10 mbar.  Diffusion of C$_2$H$_4$ from higher altitudes into this region also 
contributes to the increase.  For C$_4$H$_2$, Fig.~\ref{fignepcontc4h2} illustrates that high-altitude 
production of C$_4$H$_2$ occurs most readily during the long high-latitude summers, when the sun never sets, with 
much less production at other latitudes/times.  The seasonal and meridional behavior at 1 mbar is more complicated 
and reflects the diffusion source from higher altitudes, with its associated phase lag, as well as in situ 
production and loss.  One interesting feature that shows up in the C$_4$H$_2$ column density plot at 1 mbar is 
a brief local maximum as the sunlight returns to the high-latitude regions in the late winter or spring.

\subsection{Variations in column abundance with latitude and season on Uranus}\label{sec:columnuran}

\begin{figure*}[!htb]
%\vspace{-1.5cm}
\begin{center}
\includegraphics[clip=t,width=5.0in]{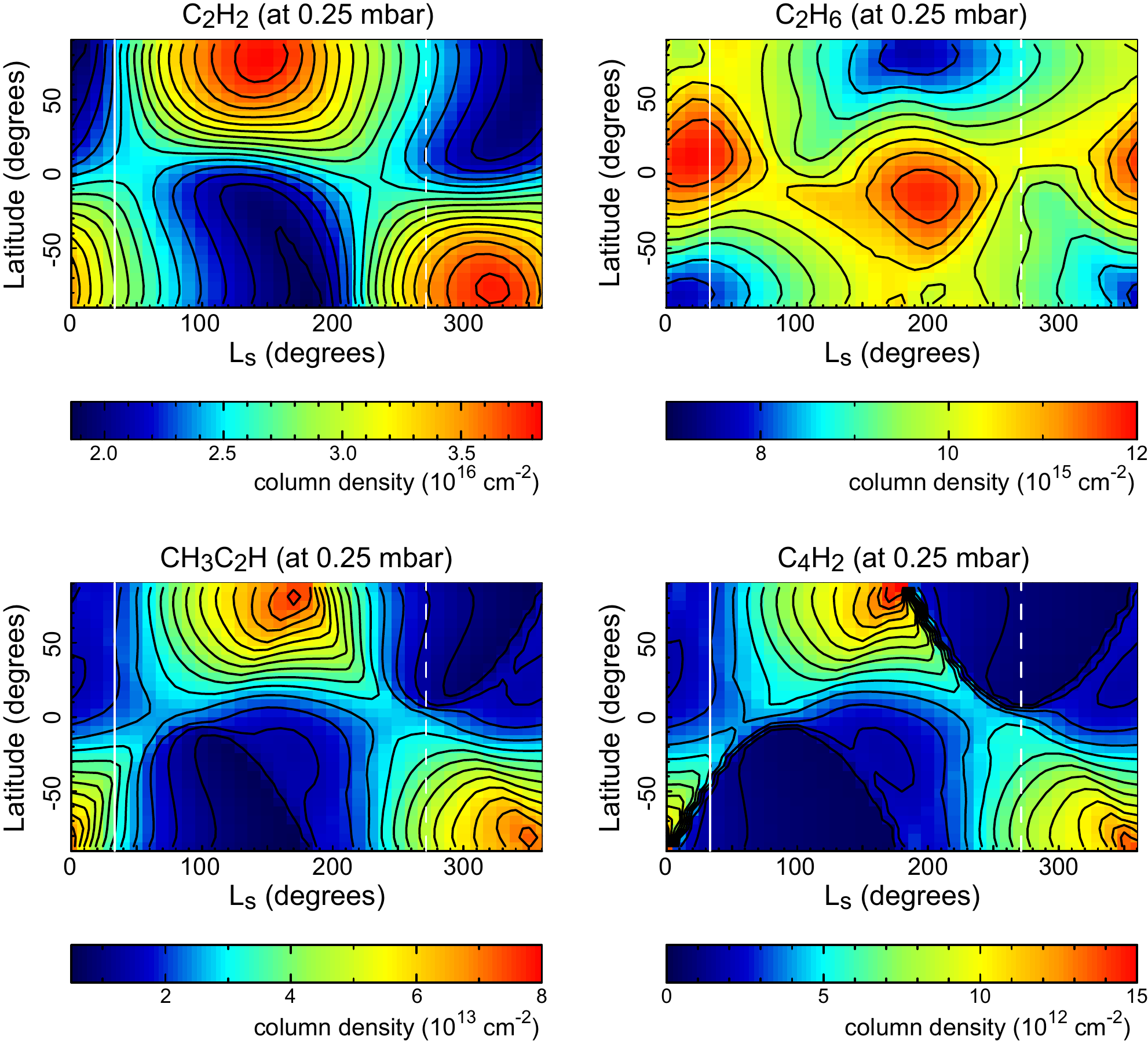}
\end{center}
\vspace{-0.5cm}
\caption{The column abundance of acetylene (Top left), ethane (Top right), methylacetylene (Bottom left), 
and diacetylene (Bottom right) on Uranus above 0.25 mbar, as a function of planetocentric latitude and 
season ($L_s$).  The dashed vertical line represents the time of the \textit{Voyager} encounter, and the 
solid vertical line represents the next Uranus opposition on 24 October 2018.
\label{figurancont}}
%\vspace{-10pt}
\end{figure*}

Figure \ref{figurancont} illustrates how the column abundances of several hydrocarbon photochemical 
products vary with latitude and season on Uranus.  Because the hydrocarbons are vertically confined to 
deeper stratospheric levels due to the weak atmospheric mixing on Uranus, we do not show the results for 
multiple pressures, but instead plot the column abundance above 0.25 mbar, which is near the peak of the 
contribution function for most of the observed hydrocarbon bands in the \textit{Spitzer}/IRS infrared 
observations \citep{orton14temp,orton14chem}.

% Fig 18, 15 lines

The column densities of C$_2$H$_2$, CH$_3$C$_2$H, and C$_4$H$_2$ exhibit maxima in the summer-to-fall 
hemispheres and minima in the winter-to-spring hemispheres, with strong hemispheric dichotomies apparent 
during most seasons.  As with Neptune, this result is caused by the greater photochemical production rates 
during periods and locations with higher mean daily solar insolation, combined with a phase lag due to the 
diffusion of hydrocarbons from higher altitudes.  Unlike Neptune, however, these species do not have a maximum 
column abundance near the equator when averaged over a year, as can be seen from the generally low column 
abundances in the equatorial region.  The abundance of diacetylene is particularly sensitive to solar 
irradiation.  Because the C$_4$H$_2$ chemical lifetime is short, the diacetylene column density drops precipitously 
in the winter polar night once the sun drops below the horizon, as is obvious in Fig.~\ref{figurancont}.  

As discussed in Section~\ref{sec:merid}, C$_2$H$_6$ is an exception to the aforementioned trends in 
the hydrocarbon distributions with latitude and season.  At 0.25 mbar and lower pressures, photolysis 
effectively destroys C$_2$H$_6$, so it survives at these pressures more readily at low latitudes and 
in the winter-to-spring hemispheres, where the mean daily solar insolation is the lowest.

\section{Comparisons with observations}\label{sec:compare}

\begin{figure*}[!htb]
%\vspace{-1.5cm}
\begin{center}
\includegraphics[clip=t,width=5.0in]{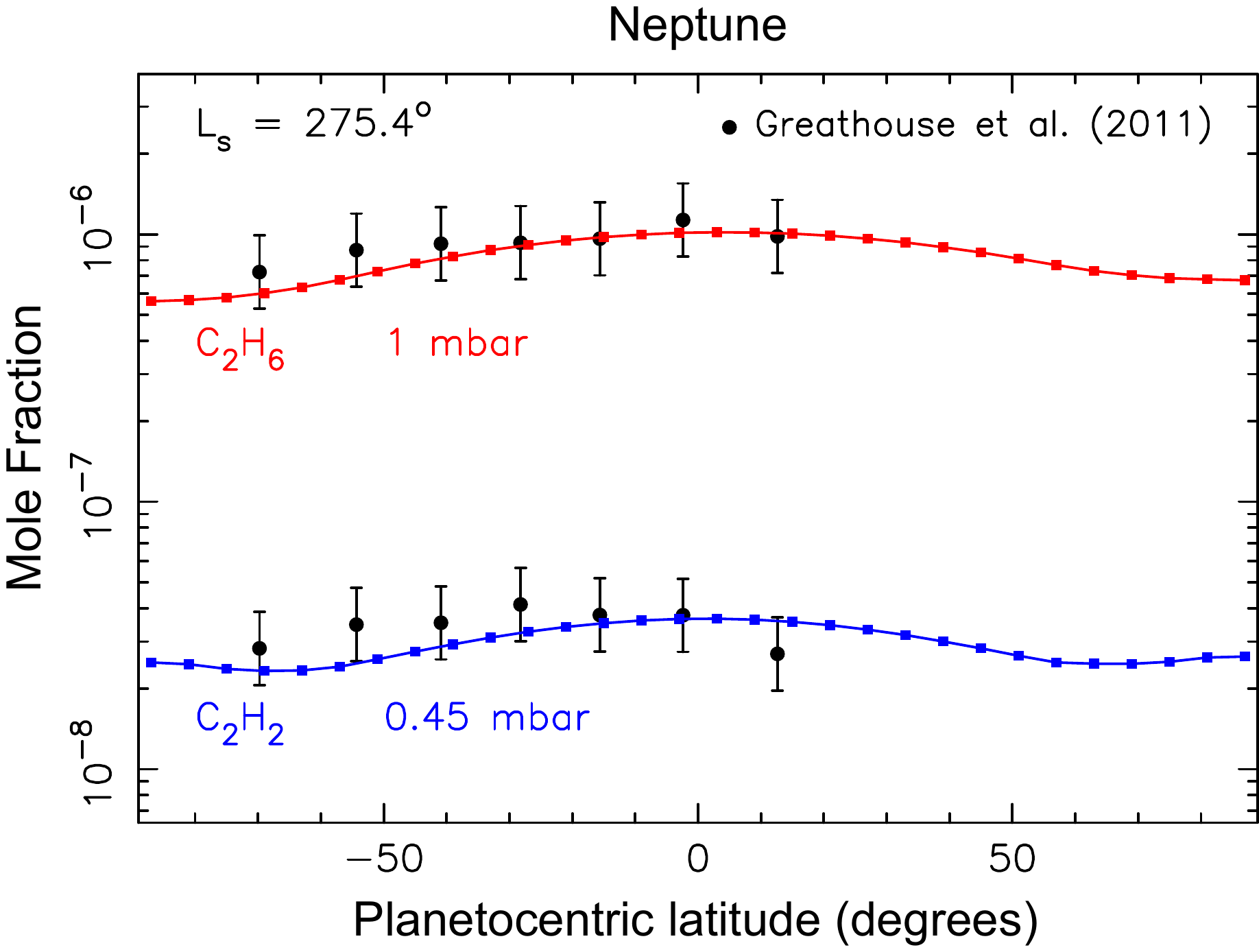}
\end{center}
\vspace{-0.5cm}
\caption{Model predictions for the meridional variation of the ethane mixing ratio at 1 mbar (red curve) and 
the acetylene mixing ratio at 0.45 mbar (blue curve) at $L_s$ = 280$\deg$, in comparison with the 
retrievals of \citet{greathouse11} from Gemini/TEXES observations at $L_s$ = 275.4$\deg$.  
\label{fignepobs}}
%\vspace{-10pt}
\end{figure*}

Low signal-to-noise ratios prevented mapping of the hydrocarbon infrared emission features 
during the \textit{Voyager} flybys of Uranus and Neptune, so no measurements of photochemical product abundances as 
a function of latitude were ever reported from the only spacecraft to ever encounter these planets.  The large 
heliocentric distances and cold atmospheric temperatures of Uranus and Neptune also pose challenges for obtaining 
spatially resolved observations of hydrocarbon photochemical products from Earth-based telescopes.  \citet{hammel07dist} 
and \citet{orton07nep} were the first to show spatially resolved images of C$_2$H$_6$ emission on Neptune, and
\citet{greathouse11} 
were the first to present meridional distributions of C$_2$H$_2$ and C$_2$H$_6$.  \citet{greathouse11} used the TEXES 
spectrograph at the Gemini North 8-m telescope in October 2007 (Neptune $L_s$ = 275.4\deg) to obtain 
high-spectral-resolution, spatially resolved observations of emission from H$_2$, CH$_4$, C$_2$H$_2$, and C$_2$H$_6$ 
at thermal-infrared wavelengths.  Emission from the S(1) rotational line of H$_2$ and the $\nu_4$ band of CH$_4$ 
were used to constrain stratospheric temperature fields, which then allowed \citet{greathouse11} to retrieve ethane 
and acetylene mixing ratios as a function of pressure and latitude.  Their results at the peak of the contribution 
functions from the strong mid-IR C$_2$H$_2$ and C$_2$H$_6$ lines are shown in Fig.~\ref{fignepobs}, in comparison with our 
results at $L_s$ = 280$\deg$ for the same pressure levels.

% Fig 19, 14 lines

Our seasonal photochemical model provides an excellent fit to the \citet{greathouse11} retrieved meridional 
distribution of C$_2$H$_6$ at 1 mbar, with both model and data exhibiting a peak mixing ratio at the 
equator that then drops off gradually toward higher latitudes in both hemispheres.  For acetylene, the observations indicate a 
slight maximum in the C$_2$H$_2$ mixing ratio at $-28\deg$ latitude, with mixing ratios dropping gradually 
toward both lower and higher latitudes.  This C$_2$H$_2$ meridional behavior is not predicted by the models, which instead 
exhibit an equatorial maximum mixing ratio with more-or-less symmetric behavior across the equator at this season (Fig.~\ref{fignepobs}), and 
a column abundance above 1 mbar that peaks at $-9\deg$ latitude (see Fig.~\ref{fignepcontc2h2}).  However, 
the C$_2$H$_2$ meridional distribution from the model, as well as an ad hoc constant-with-latitude distribution, are both 
consistent with the retrievals to within observational uncertainties.  This model-data consistency for Neptune 
hydrocarbon distributions is in sharp contrast to similar comparisons for Jupiter \citep{liang05,nixon07,nixon10,zhang13bench,fletcher16jup} 
and Saturn \citep{mosesgr05,guerlet09,guerlet10,friedson12,sinclair13,sinclair14,sylvestre15,hue15}, for which 
stratospheric circulation and/or other meridional/vertical transport processes have been suggested as influencing the 
large-scale meridional hydrocarbon distributions.  The good model-data comparisons here for Neptune suggest that either 
stratospheric winds are weaker and have less of an effect on hydrocarbon distributions on Neptune, or that 
degeneracies in retrievals and/or uncertainties in models and their related assumptions are resulting in a 
fortuitous coincidence.

Spatially resolved observations of C$_2$H$_6$ on Neptune from two different mid-infrared observational data sets 
(Keck/LWS spectra from 2003 and Gemini-S/TReCS spectra from August 2007) were also presented by \citet{fletcher14nep}.  
A factor of $\sim$2 difference in the meridionally averaged 1-mbar C$_2$H$_6$ mixing ratio between the two data sets 
highlights the difficulties inherent in calibrating mid-infrared ground-based spectra and suggests that degeneracies between 
temperatures and abundances are a perennial issue for the retrievals, particularly when dealing with low-resolution 
spectra.  \citet{fletcher14nep} also point to poor weather conditions and the lack of a Cohen standard star for 
calibration of the 2007 Gemini-S spectra.  The Fletcher et al. retrievals of the Keck data from 2003 exhibit a 
maximum in the 1-mbar C$_2$H$_6$ abundance at the equator, declining gradually toward higher latitudes --- consistent 
with both our photochemical models and the retrievals of \citet{greathouse11}.  The Fletcher et al. retrievals from 
the Gemini-S data from 2007 have a higher overall C$_2$H$_6$ mixing ratio and fine-scale meridional structure, 
including a local minimum at the equator and the south pole, that are not consistent the photochemical 
models or with the retrievals of \citet{greathouse11}.  \citet{fletcher14nep} suggest that differences between the 
2003 and 2007 data sets are caused by either poor weather for the 2007 observations, calibration issues, or by changes 
in lower-stratospheric temperatures that are not sampled reliably in retrievals of low-resolution spectra.

From the time of the \textit{Voyager} encounter with Neptune in 1989 to the 2007 observations of \citet{greathouse11} and 
\citet{fletcher14nep}, the southern hemisphere of Neptune had swung more prominently into view from the Earth, and our models 
suggest that the overall column abundance of C$_2$H$_6$ above 1 mbar would have been increasing over that time period in the southern 
hemisphere.  This trend is superficially consistent with the report of \citet{hammel06} of an increase in global-average ethane 
emission over the time period from 1985 to 2003.  On the other hand, \citet{hammel06} observe a decrease in emission 
from 2003 to 2004 that is not consistent with the photochemical models, and \citet{fletcher14nep} argue for a possible slight 
decrease in the global-average C$_2$H$_6$ mixing ratio from 2003 to mid-2007, with a sharper increase in late 2007, none of 
which are predicted by the models.  The hydrocarbon emission depends on both atmospheric temperatures and abundances, and 
it remains to be seen whether the photochemical model predictions are consistent with the observed time-variable trends and/or 
whether other factors such as adiabatic heating/cooling and variations in abundances due to transport effects are in play.

No retrievals of hydrocarbon abundances as a function of latitude have ever been published for Uranus.  The stratosphere 
of Uranus is colder than that of Neptune, and detecting any hydrocarbon emission other than from CH$_4$ and C$_2$H$_2$ from Earth-based 
telescopes has typically been difficult.  The \textit{Voyager} encounter with Uranus in 1986 occurred near southern summer 
solstice. Analyses of the ultraviolet solar occultation during the \textit{Voyager} encounter yielded C$_2$H$_2$ and C$_2$H$_6$ 
vertical profiles near the equatorial region \citep{herbert87,bishop90}.  The peak C$_2$H$_2$ mixing ratio derived from this UV 
occultation is less than that derived from global-average \textit{Infrared Space Observatory} (ISO) observations obtained 
in 1996 \citep{encrenaz98}.  Because the southern hemisphere of Uranus was still dominating the whole-disk observations in 1996, 
our photochemical model predictions for enhanced southern hemispheric C$_2$H$_2$ during this time period are qualitatively 
consistent with the greater C$_2$H$_2$ abundance seen by \citet{encrenaz98}, in comparison with the equatorial region during 
the \textit{Voyager} encounter (see Fig.~\ref{figurancont}).  

Later \textit{Spitzer}/IRS observations of Uranus acquired near 
the 2007 equinox suggest a very slight decrease in the C$_2$H$_2$ abundance in comparison with the \citet{encrenaz98} ISO 
observations.  This result, too, appears qualitatively consistent with the models, as the emission from the predicted 
greater abundance of C$_2$H$_2$ in the southern hemisphere would still be dominating the whole-disk observations at the 
equinox (Fig.~\ref{figurancont}), and the hemispheric-averaged C$_2$H$_2$ column abundance decreased between 1998 and 2007 
in the model.  The models predict a continued decrease in the C$_2$H$_2$ abundance in the southern hemisphere over the next 
few years, accompanied by an increase in the northern hemisphere abundance.

\section{Implications for future observations and modeling}\label{sec:implications}

Large ground-based telescopes are currently capable of spatially resolving Uranus and Neptune at mid-infrared wavelengths 
\citep[e.g.,][]{greathouse11,fletcher14nep,orton14workshop}, and our models provide hypotheses for the distribution 
of hydrocarbons that can be tested with such future observations.  In addition, the \textit{James Webb Space Telescope} 
(JWST), which is due to be operational shortly (currently delayed launch date March-to-June, 2019),  
%Building on the disk-averaged mid-infrared observations of Spitzer \citep{orton14temp,orton14chem}, AKARI 
%\citep{fletcher10akari}, and ISO \citep{encrenaz98}, the \textit{James Webb Space Telescope} (JWST) 
will be able to provide spatially-resolved spectroscopic maps of Uranus and Neptune.  Of particular 
interest onboard JWST is the Medium Resolution Spectrometer (MRS) of the MIRI instrument \citep{rieke15}, an Integral 
Field Unit (IFU) with the capability to provide spatially-resolved 5-28 $\mu$m spectroscopy across the planetary 
disks \citep{norwood16}.  The resulting spectral maps will provide key observational tests of the two-dimensional 
hydrocarbon distributions predicted by our models.
%, and observations of both ice giants form a part of a Guaranteed-Time Observing proposal led by H. Hammel and colleagues.

We use the time-, latitude- and altitude-dependent hydrocarbon profiles from our models to simulate the expected 
spectral radiance and brightness temperatures that could be observed by the MIRI instrument.  Our assumed spatially-uniform 
stratospheric temperatures were combined with the two-dimensional (latitude, altitude) tropospheric temperatures as 
measured by Voyager-2/IRIS \citep{orton15,fletcher14nep} to provide a realistic estimate of the temperature distribution, 
although we note that stratospheric temperature contrasts, if present, would significantly alter our results.  We employ 
the NEMESIS optimal estimation retrieval algorithm \citep{irwin08} in forward-modelling mode, simulating the 
top-of-atmosphere spectral radiance for all pixels on the observable disk, accounting for the latitude, longitude, 
and emission angle of the individual pixel element.  The spatial orientation and size of each disk was calculated for
their 2018 oppositions (3.7'' for Uranus on October 24th; 2.4'' for Neptune on September 7th).  The launch 
delay to 2019 will not have a significant effect on these simulations, given the long seasonal timescales of both 
worlds.  However, JWST is only capable of observing during a limited range of solar elongations, such that the planets 
will be observed closer to quadrature with a slightly degraded spatial resolution (angular diameter 3.6'' for Uranus, 
2.3'' for Neptune).

The MIRI forward-model simulations include both collision-induced continuum emission from H$_2$-H$_2$, H$_2$-He and 
H$_2$-CH$_4$, as well as emission and absorption from the gaseous species in our photochemical model.  Sources of spectral 
line data are described in \citet{fletcher14nep} and were used to generate $k$-distributions (absorption coefficients 
ranked in order of strength on a wavelength, temperature, and pressure grid) for use in NEMESIS.  These $k$-distributions 
were generated for CH$_4$ and its isotopologues, C$_2$H$_2$, C$_2$H$_4$, C$_2$H$_6$, C$_3$H$_4$, C$_3$H$_8$, C$_4$H$_2$, 
CO$_2$ and C$_6$H$_6$.  The tables covered each of the 12 sub-bands (i.e., four IFUs with three subbands each) of MIRI 
between 5-28 $\mu$m, using the correct variation of the spectral resolution ($R\sim1550-3250$) with wavelength.  The four 
IFUs have fields of view ranging from 3.9'' at the shortest wavelength to 7.7'' at the longest wavelength, easily 
accommodating the disks of Uranus and Neptune.  However, as the detector pixel scales (variable from 0.196'' to 0.273'') 
and image slicer scales (variable from 0.176'' to 0.645'') undersample the spectral line spread function and spatial 
point spread functions (diffraction limited spatial resolution 0.19'' at 4.89 $\mu$m to 1.10'' at 28.45 $\mu$m), respectively, 
and as the MIRI consortium is implementing dithering techniques to optimize sampling of the target field, we elected to 
base our simulation on the 0.11''/pixel plate scale of the MIRI imager, combined with the diffraction-limited performance 
of the 6.5-m mirror.  This plate scale significantly oversamples the FWHM of JWST's primary mirror and is smaller than 
the best plate scale of the MIRI IFU, but it is representative of the expected noise-free quality with an optimal use of 
telescope dithering.

Disk images at every wavelength were convolved with a Gaussian, representing the diffraction-limited spatial resolution 
of a 6.5-m primary mirror, and then averaged spectrally to highlight interesting hydrocarbon features.  Finally, a 
brightness-temperature cross section was extracted along the central meridian (to show meridional variations) and along a 
latitude circle at the sub-Earth latitude (to show the variation of brightness temperature with emission angle).  

\clearpage

% Fig 20, 21 lines
\begin{figure*}[!htb]
%\vspace{-1.5cm}
\begin{center}
\includegraphics[clip=t,width=6.0in]{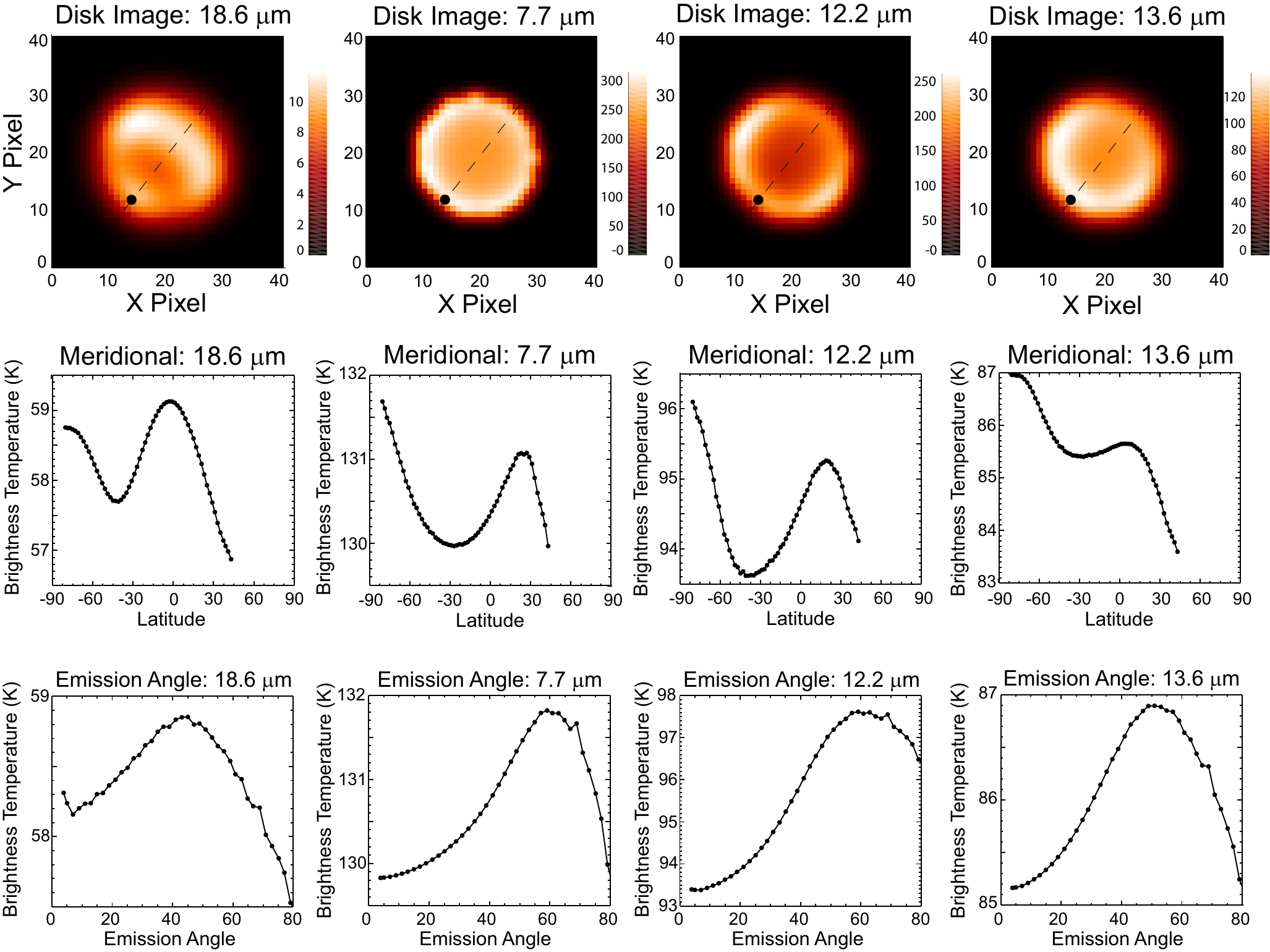}
\end{center}
\vspace{-0.5cm}
\caption{Synthetic images of Neptune (Top row) relevant to the next opposition on 7 September 2018, simulating 
what the JWST MIRI instrument would be able to see if it were operating at that time: (Left) in the tropospheric 
continuum at 18.6 $\mu$m, (Second from left) in the methane emission band at 7.7 $\mu$m, (Second from right) in 
ethane emission at 12.2 $\mu$m, and (Right) in acetylene emission at 13.6 $\mu$m.  The dashed line marks the 
central meridian, and the black dot is at the position of the south pole.  
Brightness-temperature cross sections in the same four wavelength regions were also extracted (Middle row) 
along the central meridian to highlight latitude variations and (Bottom row) along a latitude circle at the 
sub-Earth latitude to highlight variations in brightness temperature with emission angle.  
\label{figsynthnep}}
%\vspace{-10pt}
\end{figure*}

\clearpage

% Fig 21, 15 lines
\begin{figure*}[!htb]
%\vspace{-1.5cm}
\begin{center}
\includegraphics[clip=t,width=6.0in]{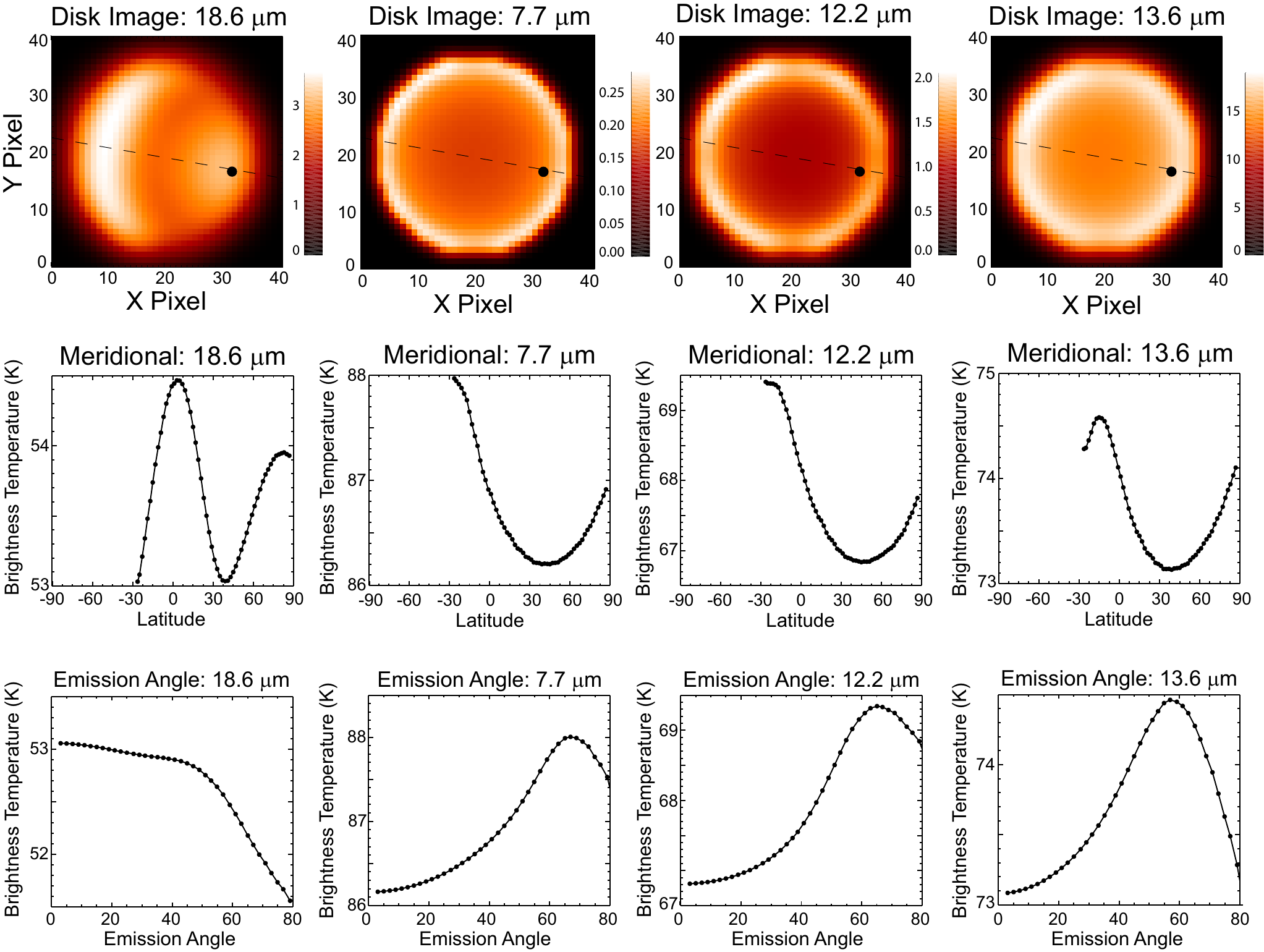}
\end{center}
\vspace{-0.5cm}
\caption{Same as Figure \ref{figsynthnep}, except for Uranus at the next opposition on 24 October 2018, 
and the black dot is at the north pole.
\label{figsynthuran}}
%\vspace{-10pt}
\end{figure*}

\clearpage

Figure \ref{figsynthnep} shows the simulated MIRI images for Neptune.  The tropospheric temperature structure is probed 
at 18.6 $\mu$m.  The assumed Voyager-era thermal structure \citep{conrath98para,fletcher14nep}, with broad mid-latitude 
temperature minima sandwiched between warmer equatorial and high-latitude regions --- which was assumed for these 
synthetic images but was not considered in the photochemical models --- shows up readily in the 18.6 $\mu$m 
images.  For a lack of information to the contrary, we have assumed a uniform stratospheric thermal structure and spatially 
uniform eddy diffusion coefficients in both the photochemical model and synthetic images, which result in roughly uniform 
methane vertical profiles across the planet.  The images in methane emission at 7.7 $\mu$m are therefore relatively bland, with limb 
brightening dominating the observed emission.  If any latitudinally variable stratospheric temperatures are actually present 
on Neptune, these should show up readily in all the hydrocarbon emission bands. Limb brightening is also apparent in the 
ethane emission at 12.2 $\mu$m and acetylene emission at 13.6 $\mu$m, but the meridional emission cross sections also show 
evidence for the predicted compositional gradients.  For example, the 1-mbar low-latitude column-density maximum in 
C$_2$H$_6$ during this season (see Fig.~\ref{fignepcontc2h6}) enhances the low-latitude emission at 12.2 $\mu$m, while the 
strong enhancement in the stratospheric column abundance on C$_2$H$_2$ at high southern latitudes is readily apparent 
in the 13.6 $\mu$m images.

Figure \ref{figsynthuran} shows the simulated MIRI images for Uranus.  As with Neptune, the Voyager-derived tropospheric 
thermal structure shows a temperature minimum at mid-latitudes \citep{flasar87,conrath98para,orton15}, which shows up 
readily in the 18.6 $\mu$m images.  The assumed uniform stratospheric temperature fields lead to rather bland images 
in the stratospheric emission features, with limb brightening dominating the emission (and note that the brightest emission 
is not actually right at the limb itself, due to an artifact of the point-spread function of the observatory, convolving dark 
space with the bright emission from the limb).  The compositional gradients are less pronounced on Uranus than on Neptune, 
so there is in general less structure apparent in the Uranus images at 12.2 and 13.6 $\mu$m.  Our models predict a 
greater column abundance of C$_2$H$_6$ at low latitudes than high latitudes during this season (see Fig.~\ref{figurancont}), 
whereas the C$_2$H$_2$ column abundance is more uniform with latitude, and hints of this behavior appear in 
the 12.2 and 13.6 $\mu$m cross sections, respectively, although any thermal contrasts in the stratosphere as a function 
of latitude would likely obscure these differences in the real images.  

It should be kept in mind that our seasonal photochemical models do not account for changes in stratospheric temperatures 
with latitude or season, for possible variations in the stratospheric methane profile with latitude or season, or for 
variations due to the advection of species.  Such effects could potentially have a major influence on the hydrocarbon 
emission profiles and vertical/meridional distributions, with meridional temperature gradients in particular having a 
likely large effect on the stratospheric emission features seen with MIRI.  
Future models should add complexity in stages, such as coupling the photochemical model with a radiative model that can accurately 
predict seasonally variable temperatures, incorporating possible meridional changes in the methane profile, and testing 2D behavior 
with different zonally average circulation scenarios.  Eventually, incorporating simple C$_2$H$_x$ chemistry and realistic 
hydrocarbon-based radiative transfer into general circulation models would be the ultimate end goal for predictions of the 
distribution of hydrocarbons as a function of latitude and season on the giant planets.

\section{Conclusions}\label{sec:conclusions}

Time-variable photochemical models are used to investigate how hydrocarbon photochemical products vary as a function of 
altitude, latitude, and season on Uranus and Neptune.  The results indicate that meridional and seasonal gradients in 
hydrocarbon abundances can persist on these planets in the absence of stratospheric circulation.

Based on our theoretical calculations and comparisons with observations, we draw the following conclusions for Neptune:

\begin{itemize}
\item Seasonal variations in hydrocarbon abundances on Neptune are predicted to be greater at high latitudes than
low latitudes because of the larger seasonal variations in solar insolation at high latitudes.
\item The larger mean daily solar insolation in the summer hemisphere leads to greater photochemical production and 
higher abundances of hydrocarbon photochemical products in the summer hemisphere than the winter hemisphere.
\item Seasonal variations in hydrocarbon abundances on Neptune are most pronounced at high altitudes and become 
progressively weaker at depth because diffusion time constants and chemical lifetimes increase with increasing pressure.  
Hydrocarbons respond quickly to changes in the solar flux at high altitudes, but the 
greater time constants at depth introduce phase lags in the seasonal response to the changing insolation, such that 
the abundance maxima for long-lived hydrocarbons such as C$_2$H$_2$ and C$_2$H$_6$ shift to later in the summer season 
with increasing depth.  
\item In the absence of advection or other stratospheric transport processes, meridional variations in hydrocarbon abundances 
will exist on Neptune and will again be greater at high altitudes than low altitudes.  At pressures greater 
than a few mbar, chemical and transport time scales are longer than a Neptune season, and the yearly averaged solar 
flux then controls the hydrocarbon abundance variation with latitude.  Because the yearly average insolation is larger 
at the equator than the poles at Neptune, most hydrocarbons at pressures greater than a mbar are 
expected to exhibit abundance maxima at low latitudes, decreasing gradually toward higher latitudes.  Hydrocarbons 
that have short photochemical lifetimes at mbar pressures, such as C$_2$H$_4$ and C$_4$H$_2$, exhibit additional 
complicated seasonal behavior.
\item Model predictions for the meridional variation of C$_2$H$_2$ and C$_2$H$_6$ on Neptune compare well with 
retrievals from the spatially resolved Gemini infrared spectral observations of \citet{greathouse11}, and the Keck 
observations of \citet{fletcher14nep} (to within uncertainties of the observational analyses), suggesting that 
stratospheric circulation is less important in controlling the large-scale hydrocarbon meridional distributions on
Neptune than it is on Jupiter and Saturn 
\citep[cf.][]{liang05,mosesgr05,nixon07,guerlet09,sinclair13,sylvestre15,hue15,fletcher16jup}.
\end{itemize}

Neptune's axial tilt is only a couple degrees greater than that of Saturn, and the predicted seasonal variations on 
Neptune share many similarities with those on Saturn \citep[cf.][]{mosesgr05,hue15}.  Seasonal variations on Uranus, 
on the other hand, differ notably from both Saturn and Neptune.  Although the extreme axial tilt of Uranus is partially 
responsible for these differences, the very weak atmospheric mixing on Uranus plays a larger role.
Our main conclusions with respect to Uranus are the following:

\begin{itemize}
\item The weak vertical transport on Uranus confines hydrocarbons to relatively low altitudes.  The vertical diffusion 
time constants and chemical lifetimes tend to be large in the pressure region where the complex hydrocarbons are being produced, 
which results in more muted seasonal variations on Uranus than on Neptune.  Some seasonal variation is expected in the 
0.1-1 mbar region of Uranus (leading to notable hemispheric dichotomies in hydrocarbon abundances, with generally greater abundances 
in the summer-to-fall hemisphere than the winter-to-spring hemisphere), but virtually no seasonal changes are predicted at 
pressures greater than 1 mbar.
\item Most hydrocarbons on Uranus are predicted to have a greater abundance at the poles than the equator, due to the 
latitude variation of the annual average solar insolation on the highly tilted Uranus.  Ethane is an exception to this 
trend because of effective photolysis loss that can compete with production; C$_2$H$_6$ abundances are on average greater 
at low latitudes than high latitudes.
\end{itemize}

Solar Lyman alpha photons scattered from atomic hydrogen in the interplanetary medium provide an important source 
of photolyzing radiation for both Uranus and Neptune, particularly in the high-latitude polar winter, where sunlight 
is absent for long periods of time each year.  Seasonal variations at high altitudes would be even more significant 
without this extra UV source.

Based on our photochemical model results, we simulated the emission from Uranus and Neptune that would be observed
using the MIRI instrument onboard JWST.  We find that hydrocarbon variations with latitude should show up readily in 
MIRI images of Neptune but would be more muted for Uranus.  Keep in mind, however, that the models presented here 
do not consider possible seasonal variations in stratospheric temperatures, 
possible latitude variations in the stratospheric methane abundance or its vertical profile, or possible perturbations 
due to stratospheric circulation, waves, or other transport processes.  These factors could potentially play a major role 
in controlling large-scale vertical and meridional hydrocarbon distributions, particularly on Uranus, where chemical and diffusion 
time constants are long.  Because many of the hydrocarbon photochemical products condense in the cold lower stratospheres 
of these planets, any (likely) changes in temperature with season could strongly affect the total column density of the 
condensable products, due to the high sensitivity of the species' vapor pressures to temperatures, although the gas-phase 
chemistry variations themselves are not very sensitive to temperature \citep[e.g.,][]{mosesgr05,moses15}.  
Differences in the methane abundance at high versus low latitudes have already been identified in the tropospheres of 
Uranus and Neptune \citep{karkoschka09,karkoschka11,sromovsky11,sromovsky14,irwin12,tice13,dekleer15,luszczcook16}, and 
if such differences extend into the stratosphere, would have a major effect on photochemical production rates.  Seasonal 
forcing likely drives stratospheric circulation on Uranus and Neptune \cite[e.g.,][]{flasar87,conrath90,conrath91nep}, 
and could have important consequences for hydrocarbon distributions.  The models presented here represent a first-order 
solution that when compared to observations can provide valuable insights into the physical and chemical processes at 
play in the stratospheres of Uranus and Neptune.  Future 1D, 2D, and 3D models could add complexity, as needed, to explain 
observations. 
We present our full model results in the journal supplementary material in the hopes that they will be of use in 
analyzing future observations, including potential spatially resolved infrared spectral observations from JWST 
\citep[e.g.,][]{norwood16} and potential future spacecraft missions to the ice giants \citep[e.g.,][]{hofstadter17,mousis17}.

Hydrocarbon photochemical products (both vapor and aerosols) on Uranus and Neptune help control stratospheric heating, 
cooling, and energy balance, which in turn influence atmospheric dynamics.  The gas-phase hydrocarbons reveal the complex 
chemistry at play in these atmospheres and can act as tracers to illuminate dynamical motions.  Furthering our understanding 
of the three-dimensional time-variable behavior of hydrocarbons on Uranus and Neptune is therefore important for furthering 
our understanding of the complex chemical, radiative, and dynamical couplings and feedbacks that characterize our 
solar-system ice giants.

\section{Acknowledgments}\label{sec:acknow}

This material is based on research supported by the National Aeronautics and Space Administration (NASA) Science 
Mission Directorate under grant NNX13AH81G from the Planetary Atmospheres Research Program.  The oxygen chemistry 
portion was supported by NASA grant NNX13AG55G.  Fletcher was supported by a Royal Society Research Fellowship 
and European Research Council Consolidator Grant (under the European Union's Horizon 2020 research and innovation 
programme, grant agreement No.~723890) at the University of Leicester.  Orton acknowledges support from 
NASA to the Jet Propulsion Laboratory, California Institute of Technology.



\bibliographystyle{elsarticle-harv} 
\bibliography{references}


%\begin{thebibliography}{00}



%\end{thebibliography}

\end{document}